\pdfoutput=1
\documentclass[12pt]{iopart}

\expandafter\let\csname equation*\endcsname\relax
\expandafter\let\csname endequation*\endcsname\relax
\usepackage[perpage,marginal]{footmisc}
\usepackage[unicode, pdfusetitle]{hyperref}
\usepackage[all]{hypcap}
\usepackage{amssymb,amsmath,verbatim,mathtools,needspace,enumitem,etoolbox,graphicx,lmodern,microtype}
\usepackage[usenames,dvipsnames]{xcolor}
\usepackage[T1]{fontenc}
\usepackage[utf8]{inputenc}
\usepackage{pifont}
\usepackage{yfonts}
\usepackage{url}
\usepackage[normalem]{ulem}
\definecolor{linkcolor}{rgb}{0.0,0.3,0.5}
\hypersetup{colorlinks=true,linkcolor=linkcolor,citecolor=linkcolor,filecolor=linkcolor,urlcolor=linkcolor}

\usepackage{xspace}
\usepackage{bm}
\usepackage[numbers,square,comma,sort&compress]{natbib}

\usepackage[bottom=2.5cm]{geometry}

\newcommand{\caltech}{{TAPIR 350-17, California Institute of Technology, 1200 E California Boulevard, Pasadena, CA 91125, USA}}

\newcommand{\olemiss}{{Department of Physics and Astronomy, The University of Mississippi, University, MS 38677, USA}}

\begin{document}

\title[The binary black hole explorer: on-the-fly visualizations\ldots]{The
    binary black hole explorer: on-the-fly visualizations of precessing binary
black holes}

\hypersetup{pdfauthor={Varma, Stein, and Gerosa}}

\author{
Vijay~Varma$^{1}$,
Leo~C.~Stein$^{2}$,
Davide~Gerosa$^{*\,1}$
}
\address{$^{1}$~\caltech}
\address{$^{2}$~\olemiss}

\address{*~Einstein Fellow}

\ead{\href{mailto:vvarma@caltech.edu}{vvarma@caltech.edu}}

\graphicspath{{Figs/}}
\newcommand{\roughly}{\mathchar"5218\relax} %

\newcommand{\red}{\textcolor{red}}
\newcommand{\vv}[1]{\textcolor{WildStrawberry}{#1}}
\newcommand{\lcs}[1]{\textcolor{Cerulean}{LCS: #1}}
\newcommand{\dg}[1]{\textcolor{blue}{#1}}

\newcommand{\Note}[1]{\textcolor{green}{\textbf{[#1]}}}
\newcommand{\h}{\mathpzc{h}}
\newcommand{\hlm}{\mathpzc{h}_{\ell m}}
\newcommand{\Alm}{A_{\ell m}}
\newcommand{\omegalm}{\omega_{\ell m}}
\newcommand{\philm}{\phi_{\ell m}}
\newcommand{\chieff}{\chi_{\mathrm{eff}}}

\newcommand{\ts}{\mathcal{TS}}
\newcommand{\ps}{\mathcal{PS}}
\newcommand{\bchi}{\bm{\chi}}

\newcommand{\pd}{\partial}

\newcommand{\PackageName}{\emph{binaryBHexp}\xspace}

\begin{abstract}
Binary black hole mergers are of great interest to the astrophysics community,
not least because of their promise to test general relativity in the highly
dynamic, strong field regime. Detections of gravitational waves from these
sources by LIGO and Virgo have garnered widespread media and public attention.
Among these sources, precessing systems (with misaligned black-hole
spin/orbital angular momentum) are of particular interest because of the rich
dynamics they offer. However, these systems are, in turn, more complex compared
to nonprecessing systems, making them harder to model or develop intuition
about. Visualizations of numerical simulations of precessing systems provide a
means to understand and gain insights about these systems. However, since these
simulations are very expensive, they can only be performed at a small number of
points in parameter space. We present \emph{binaryBHexp}, a tool that makes use
of surrogate models of numerical simulations to generate on-the-fly interactive
visualizations of precessing binary black holes. These visualizations can be
generated in a few seconds, and at any point in the 7-dimensional parameter
space of the underlying surrogate models. With illustrative examples, we
demonstrate how this tool can be used to learn about precessing binary black
hole systems.
\end{abstract}

\section{Introduction}
\label{sec:intro}

The merger of two black holes (BHs) is one of the most violent events in the
Universe.  In the span of a few seconds, the incredible amount of energy
$\roughly 10^{60}$MeV~\cite{LIGOVirgo2016a} 
 is liberated in gravitational waves (GWs).  These ``ripples in
spacetime'' travel across the Universe at the speed of light to our detectors,
providing us unique insights into these spectacular astrophysical events.

The first direct detection~\cite{LIGOVirgo2016a} of GWs from a BH merger was
achieved in 2015 by the LIGO~\cite{aLIGO2} twin detectors. This is one of the
greatest achievements in modern science, crowning decades of theoretical and
experimental efforts in gravitational physics. The detection of GWs
not only earned the 2017 Nobel Prize in physics~\cite{NobelPrize2017}, but also
sparked an unprecedented interest in science among the general public. For a
few days, BHs were on the front pages of most newspapers in the world!

Despite the immense technical difficulties in detecting them, astrophysical BHs
are remarkably simple objects, characterized only by their mass and spin. From
far away they can be thought of as the analogs of Newtonian point masses in
Einstein's general relativity (GR). Near a BH, departures from Newtonian
gravity such as the event horizon, gravitational lensing, gravitational time
dilation, frame dragging, etc, become apparent.  

When in a binary system, the departure is even more drastic. First, there are
no stable binary orbits in GR: emission of GWs takes away energy, angular
momentum, and linear momentum from the system, causing the binary's orbit to
shrink. Second, in Newtonian gravity, a point-mass binary orbit that starts in
the equatorial plane remains in the equatorial plane. In GR, on the other hand,
if the BH spins are misaligned with respect to the orbital angular momentum,
relativistic spin-spin and spin-orbit couplings cause the system to
precess~\cite{Apostolatos1994,Kidder:1995zr,Racine:2008qv,Gerosa:2015tea}.
Much like a top whose spin axis is misaligned with the orbital angular
momentum, the spins and the orbital angular momentum oscillate about the
direction of the total angular momentum. This precession is imprinted on the
observed gravitational waves as characteristic modulations of amplitude and
frequency.

The evolution of a binary BH system can be divided into three stages: inspiral,
merger, and ringdown. During the inspiral, the BHs gradually approach each
other due to loss of energy and angular momentum to GWs. As they get closer,
they eventually coalesce and merge. After the merger, one is left with a
single, but highly distorted, BH.  In the final stage, called ringdown, all
these perturbations (``hairs'') are radiated away and the remnant settles down
to its final steady state. The remnant BH is characterized entirely by it mass,
spin, and recoil velocity (or ``kick''). These properties are associated with
the asymptotic conservation laws of energy, angular momentum, and linear
momentum, respectively.

\begin{figure*}[h]
\makebox[\linewidth][c]{%
\includegraphics[scale=0.6]{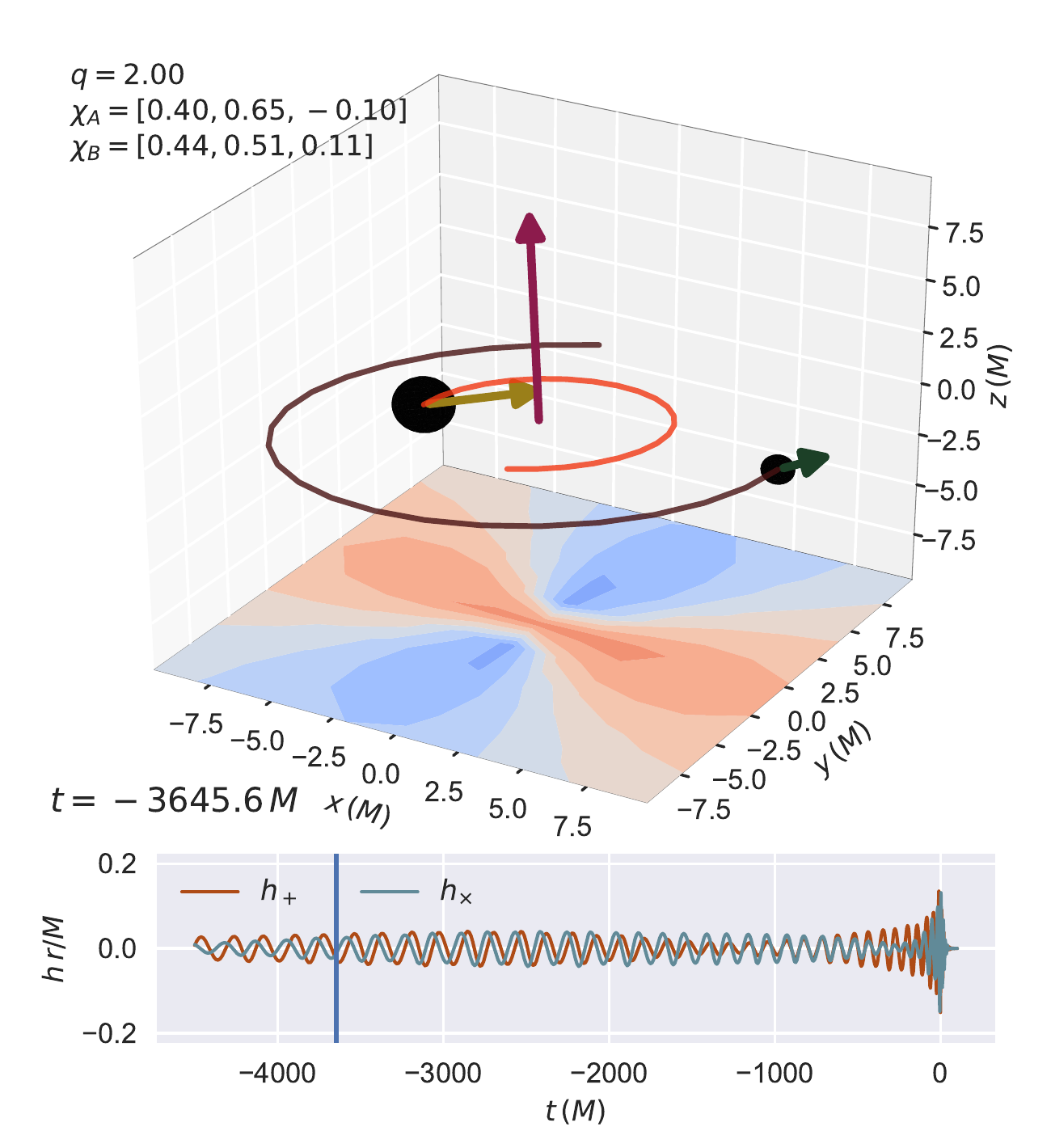}
\includegraphics[scale=0.6]{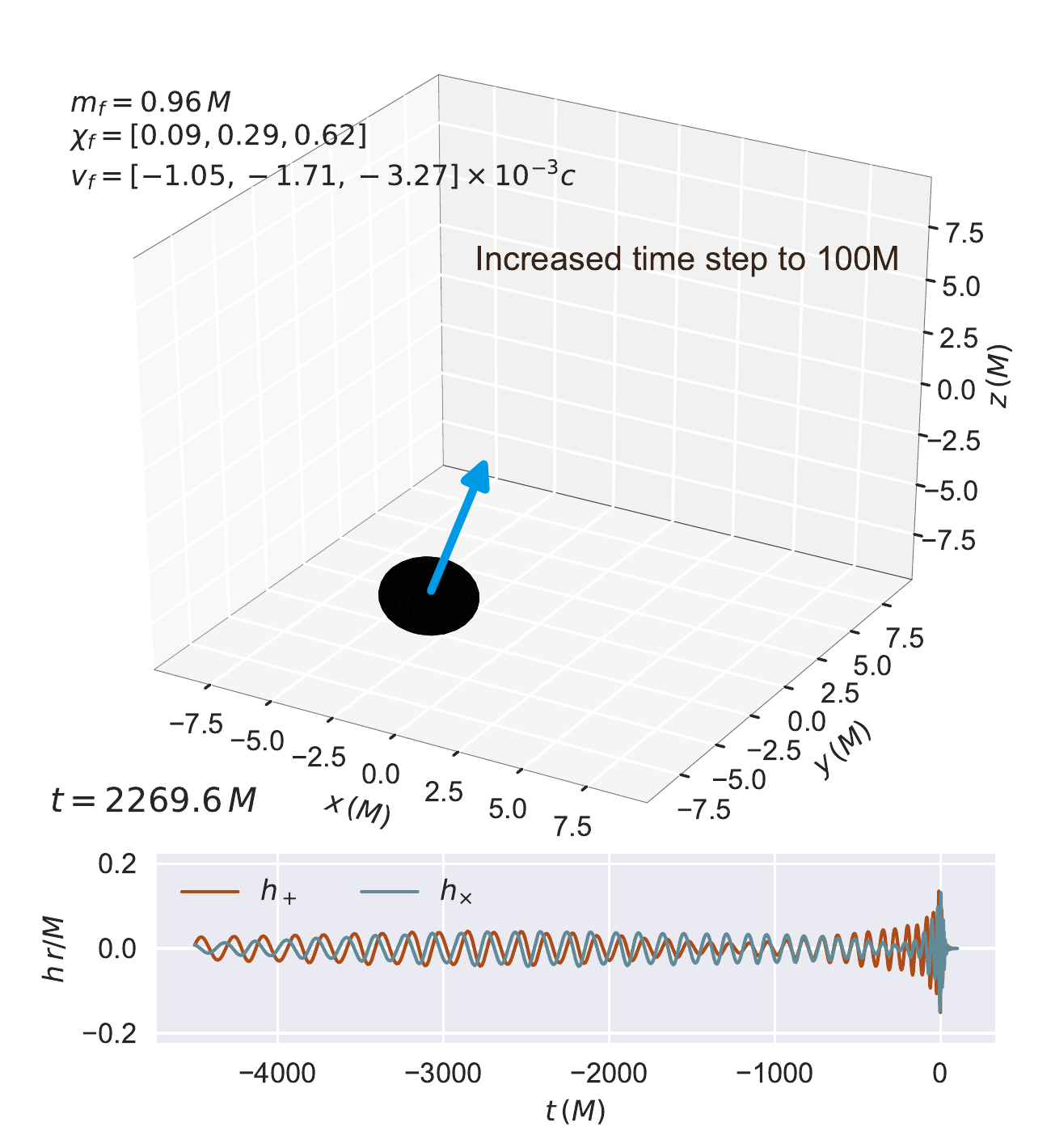}
}
\makebox[\linewidth][c]{%
\includegraphics[scale=0.19]{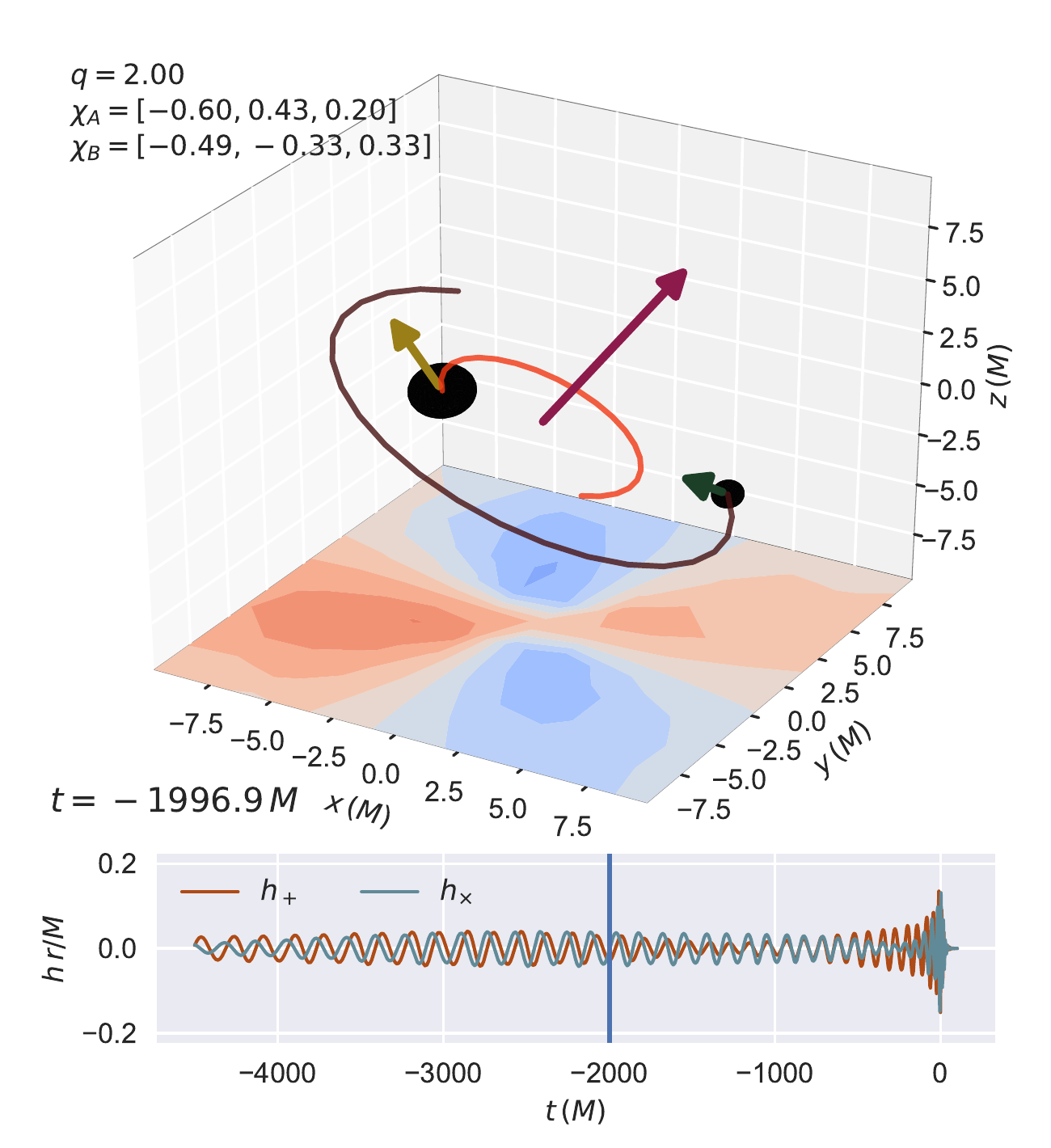}
\includegraphics[scale=0.19]{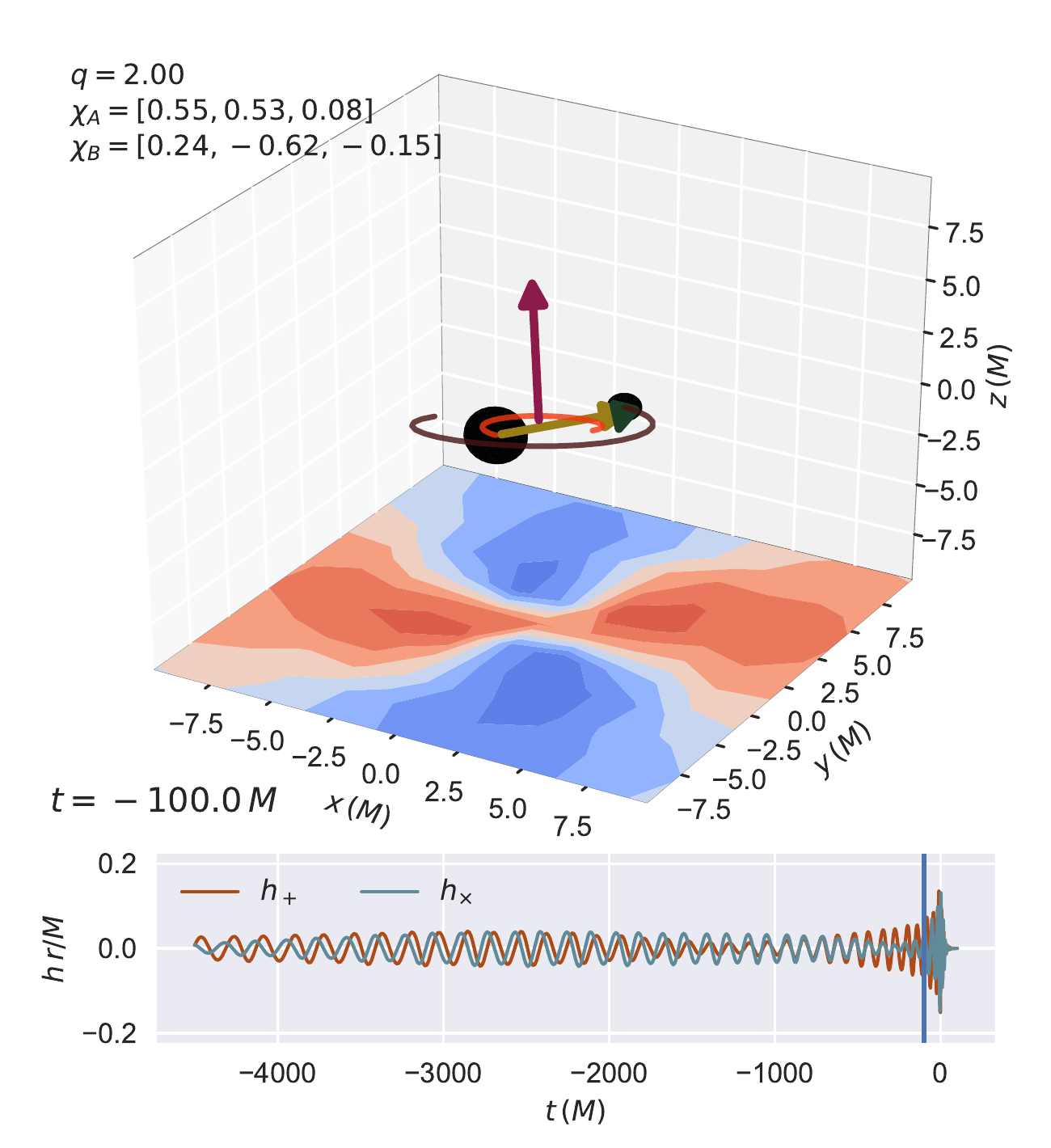}
\includegraphics[scale=0.19]{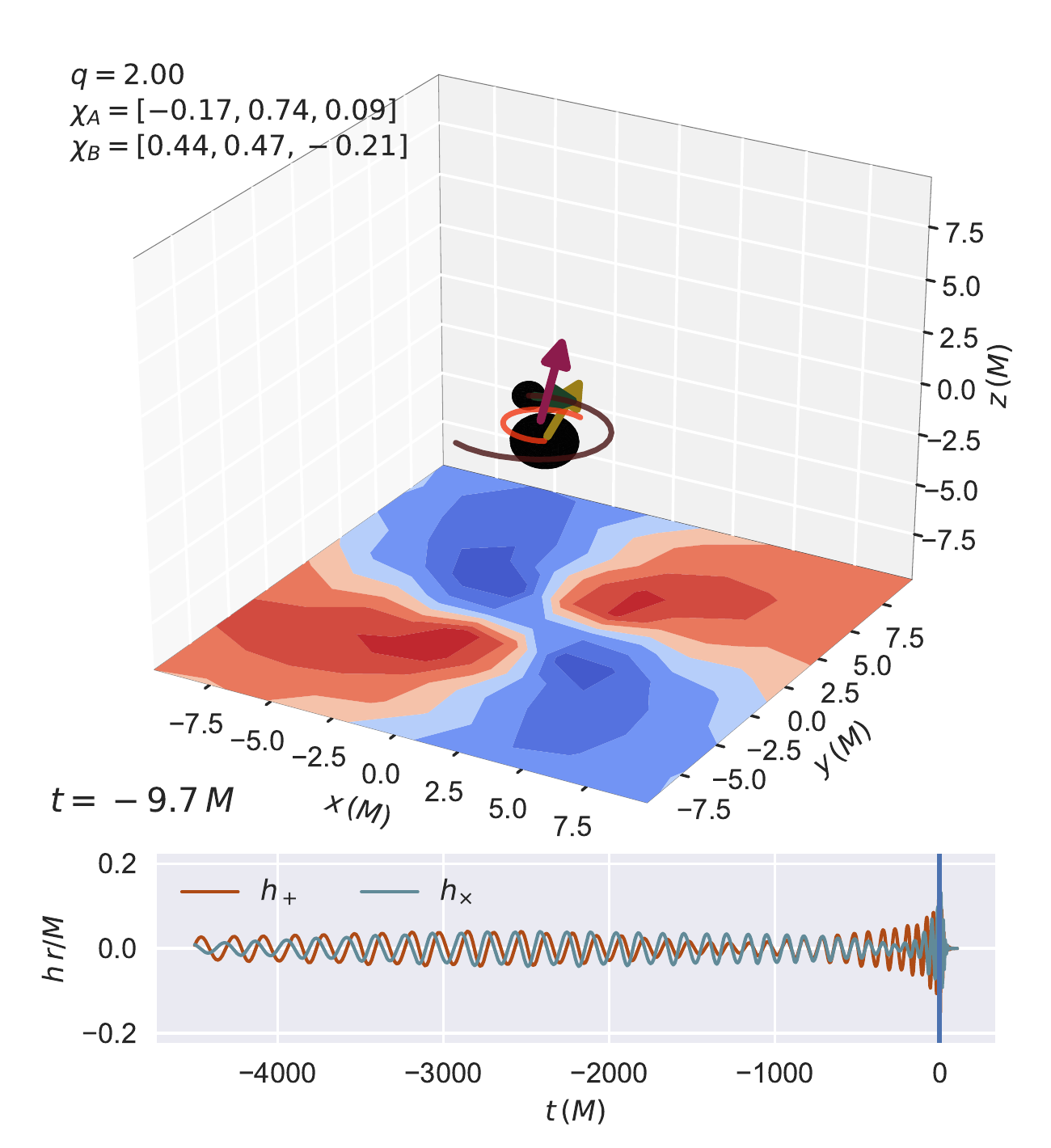}
\includegraphics[scale=0.19]{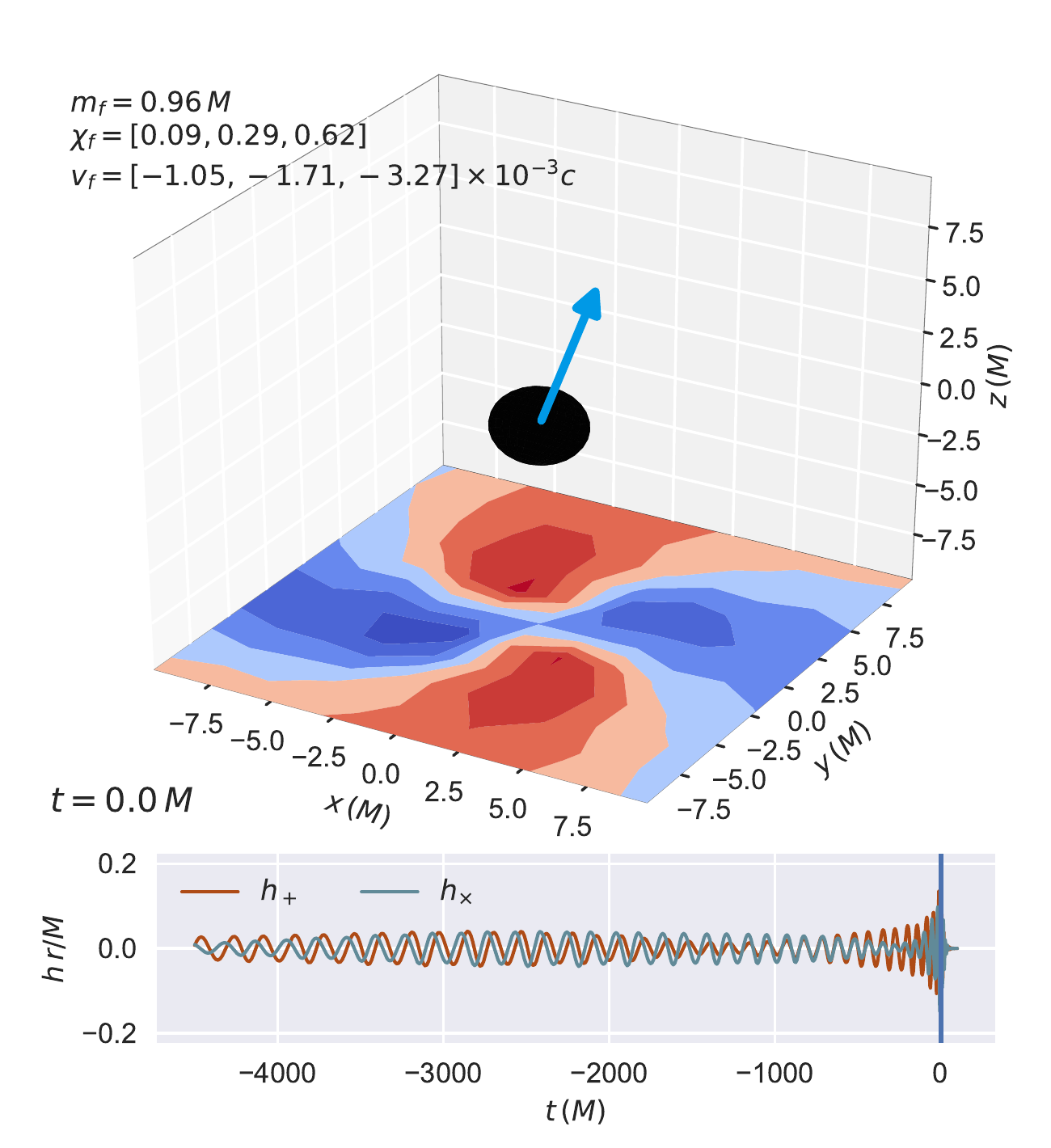}
\includegraphics[scale=0.19]{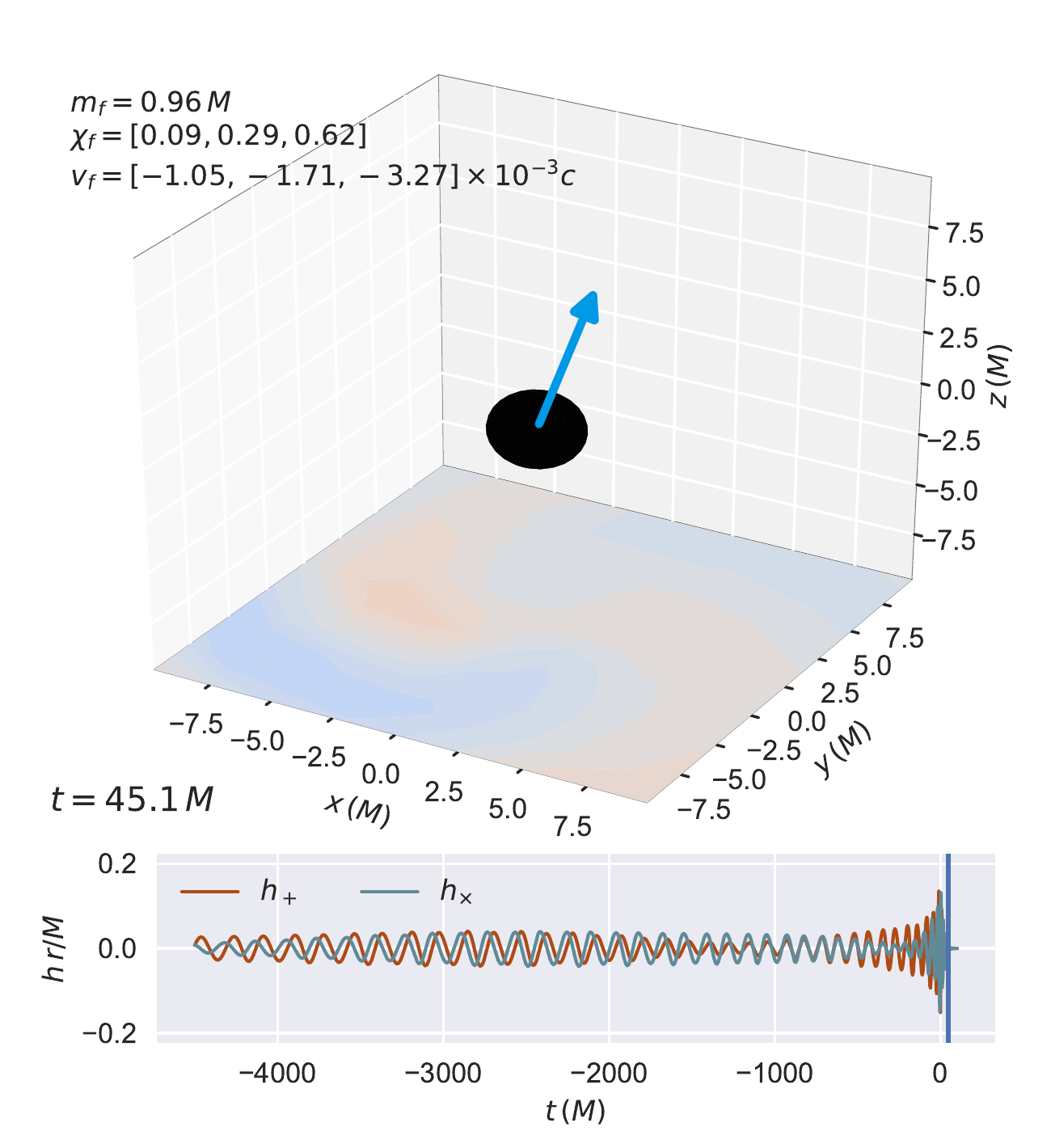}
\includegraphics[scale=0.19]{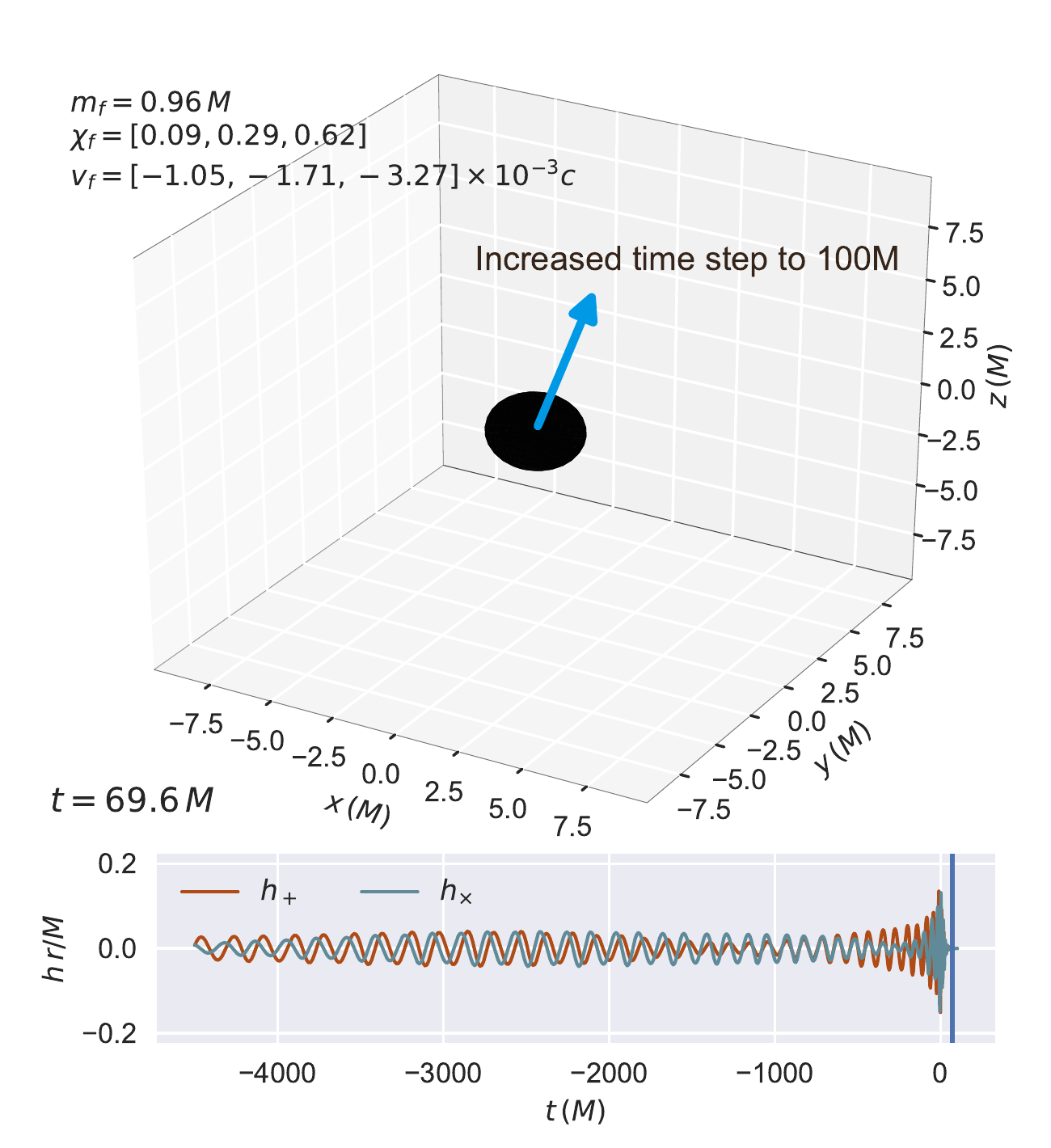}
}
\caption{Snapshots during the inspiral (top-left), post-ringdown (top-right),
and intermediate (bottom) stages of a precessing binary BH evolution. Each BH
horizon is represented by an oblate spheroid. The arrows on the BHs indicate
the spin vectors; the larger the spin the longer the arrow. The arrow centered
at the origin indicates the orbital angular momentum. On the bottom plane, we
show the plus polarization of GWs, as seen by an observer at each point.  Red
(blue) colors indicate positive (negative) values. Notice the quadrupolar
nature of the emitted waves. The subplots at the bottom of each panel show GW
plus and cross polarizations, as seen by a far-away observer viewing from the
camera viewing angle. The time to the peak of the waveform amplitude is
indicated in the figure text as well as the slider in the bottom subplots.
This animation is available at
\href{https://vijayvarma392.github.io/binaryBHexp/\#prec_bbh}{vijayvarma392.github.io/binaryBHexp/\#prec\_bbh}.
}
\label{fig:visualization_example}
\end{figure*}

Modeling GWs emitted during all three stages is crucial to interpreting
observations from detectors like LIGO~\cite{aLIGO2} and
Virgo~\cite{TheVirgo:2014hva}. The merger phase, in particular, can only be
captured accurately with expensive numerical-relativity (NR) simulations (see
e.g. Ref.~\cite{Lehner:2014asa} for a review). Obtaining a single merger
waveform prediction might take months of computational time on powerful
supercomputers.  Visualizations~\cite{youtubeAnimations} of these simulations
have been instrumental in disseminating GW discoveries for outreach and
educational purposes. To some extent, experts in the field also rely on visual
products to develop intuition and illuminate future directions for research. In
particular, visualizations of precessing binary BHs can give valuable insights
into their complex dynamics.  Available visualizations directly rely on NR
simulations, and are therefore restricted to the small number of configurations
which have been simulated.  Generating a new visualization at a generic point
in parameter space would involve a new, expensive NR simulation.

In this paper, we present the ``binary Black Hole explorer'' (\PackageName): a
new tool to generate on-the-fly, yet accurate, interactive visualizations of
precessing binary black hole evolutions with arbitrary parameters. We rely on
recent NR surrogate models. Trained against several hundreds of numerical
simulations, these models have been shown to accurately model both the emitted
gravitational waveform~\cite{Blackman:2017pcm} and the BH remnant
properties~\cite{Varma:2018aht} of precessing binary BH systems. With our
easy-to-install-and-use Python package, one can generate visualizations within
a few seconds on a standard, off-the-shelf, laptop computer. Some examples are
available at
\href{https://vijayvarma392.github.io/binaryBHexp}{vijayvarma392.github.io/binaryBHexp}.

Figure~\ref{fig:visualization_example} shows snapshots from a visualization
generated with \PackageName. During the inspiral, both radiation reaction and
spin precession are at play.  While the separation shrinks because of GW
emission, the orientations of the spins, and the orbital angular momentum, all
vary in time. The GW emission frequency gradually scales as $f \sim
r_{12}^{-3/2}$, and amplitude scales as $h\sim r_{12}^{-1}$, where $r_{12}$ is
the binary separation, producing a distinctive ``chirp'' where both frequency
and amplitude sweep up over time.  GWs are emitted in two polarizations, $h_+$
and $h_\times$, as predicted by Einstein's GR. As explored later, the relative
amplitude of the two polarizations crucially depends on orientation of the
observer with respect to the binary. Spin precession causes amplitude
modulations during the inspiral phase, which are also dependent on the observer
orientation.  After merger, the component BHs are replaced by a remnant BH,
whose properties are determined by conservation laws, as mentioned above. The
merger process emits copious gravitational radiation, and corresponds to the
peak amplitude of the waveform.

The rest of the paper is organized as follows. Sec.~\ref{sec:methods} describes
methods and approximations employed to generate visualizations such as
Fig.~\ref{fig:visualization_example}.  In Sec.~\ref{sec:explorations}, we
demonstrate the power of this tool with several examples aimed at exploring
known phenomenology in BH dynamics. Sec.~\ref{sec:python_implementation}
describes code implementation and usage. Finally, we provide concluding remarks
in Sec.~\ref{sec:conclusion}.

\section{Methods}
\label{sec:methods}

\subsection{Preliminaries}

\begin{figure*}
\begin{center}
\includegraphics[scale=1]{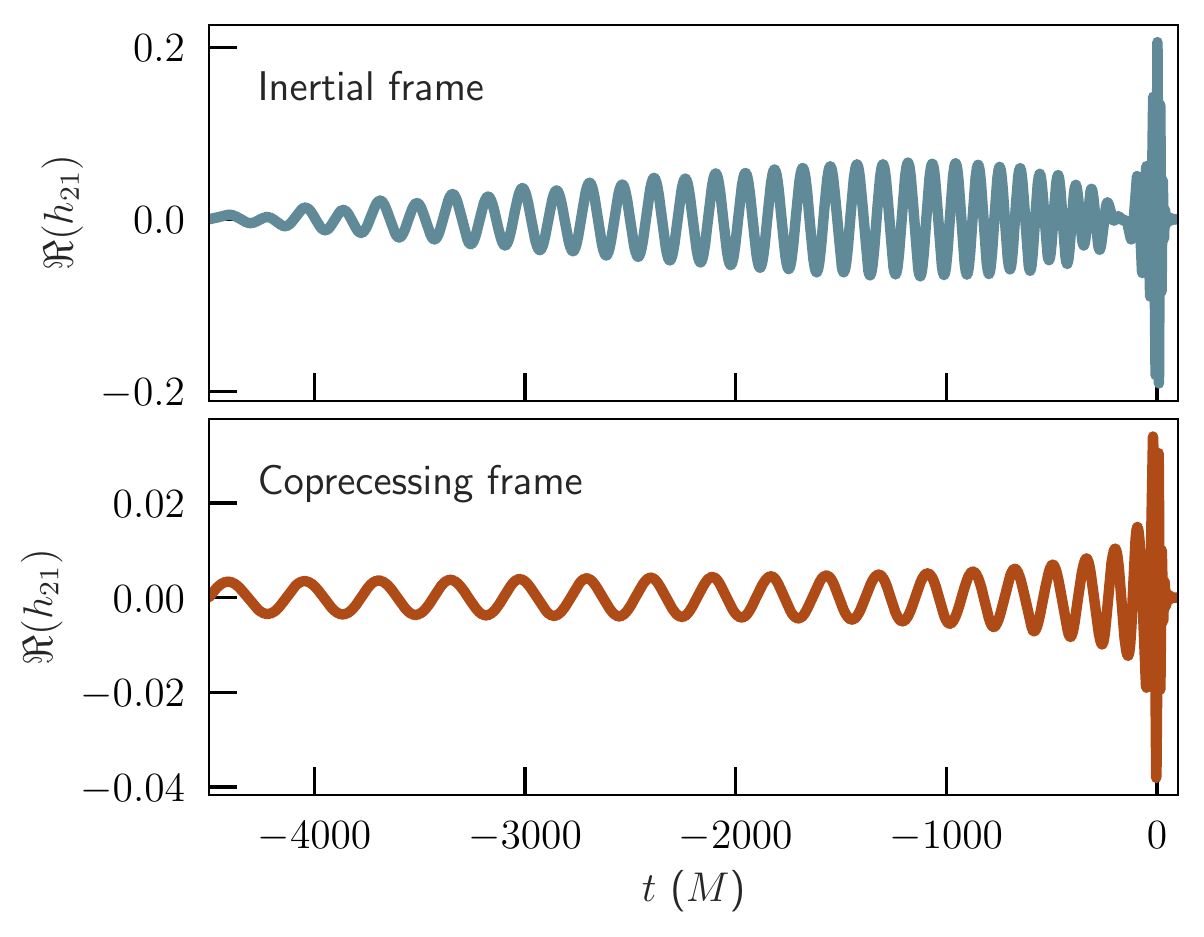}
\end{center}
\vspace{-1em}
\caption{Example of the real part of the $(\ell=2,m=1)$ spin-weighted
spherical harmonic mode (see Sec.~\ref{subsec:gw_methods}) of the GW for a
precessing black hole binary, in the inertial (top) and coprecessing (bottom)
frames. $t=0$ corresponds to the peak of the waveform amplitude.
\vspace{-1em}
}
\label{fig:precessing_frames}
\end{figure*}

We start with some definitions, referring the reader to standard GR and GW
textbooks for more details~\cite{schutz:aFirstCourseInGR, Maggiore2008,
baumgarteShapiroBook, hartle:gravity, carrollTextbook, MTW}. Throughout this
paper, we use geometric units with $G=c=1$.

An isolated astrophysical BH is characterized entirely by its mass $m$ and spin
angular momentum $\mathbf{S}=\bm{\chi} m^2$. $\bm{\chi}$ is the dimensionless
spin, with magnitude $\chi\leq1$,  and $\bm{a}=\bm{\chi} m$ is the Kerr
parameter.

A quasicircular precessing binary BH system is characterized by seven intrinsic
parameters: mass ratio $q=m_1/m_2$, and two spin vectors $\bm{\chi}_1$,
$\bm{\chi}_2$. Here, subscript 1 (2) corresponds to the heavier (lighter) of
the two BHs. The total mass of the system $M=m_1+m_2$ can be scaled out.
Therefore, throughout this paper, all length and time quantities are in units
of $M$. Similarly, all frequency quantities are in units of $1/M$. After the
merger takes place, the remnant BH is characterized by its mass $m_f$, spin
$\bm{\chi}_f$ and recoil velocity $\bm{v}_f$.

If the BH spins are \mbox{(anti-)aligned} with respect to the orbital angular
momentum $\bm{L}$, the emitted GWs have monotonically increasing amplitude
and frequency. Instead, if the component spins are misaligned with respect to
$\bm{L}$, couplings between the momenta $\bm{L}$, $\bm{S}_1$, and $\bm{S}_2$
cause them to precess about the direction of the total angular momentum
$\bm{J}=\bm{L}+\bm{S}_1+\bm{S}_2$. GW amplitude and frequency are not
monotonic, and their modulations strongly depend on the viewing angle
\cite{Apostolatos1994}.  This complexity can be in part removed by moving into
a non-inertial reference frame which tracks the direction of
$\bm{L}$~\cite{Boyle:2011gg, Schmidt2010, OShaughnessy2011}. In this
\emph{coprecessing} frame, the waveform looks nearly as simple as that of a
nonprecessing source (cf.\ bottom panel of Fig.~\ref{fig:precessing_frames}),
and can be modeled with methods developed to study nonprecessing systems.

\subsection{Surrogate models}

NR surrogate models provide a fast-but-accurate method to model GW signals. We
use a model developed by Blackman et al. \cite{Blackman:2017pcm} named
NRSur7dq2 to predict both the waveform and the BH spin dynamics.  NRSur7dq2 was
trained against 886 NR simulations in the 7-dimensional parameter space of mass
ratios $q \leq 2$, and dimensionless spin magnitudes $\chi_1,\chi_2\leq0.8$.
NRSur7dq2 predicts both the emitted GWs and the associated BH spin dynamics. In
particular, it models four important quantities that we make use of in this
work: (i) the waveform modes $h_{\ell m}$ expanded in spin-weighted spherical
harmonics (cf.\ Sec.~\ref{subsec:gw_methods}); (ii) the unit quaternions
$\hat{Q}(t)$ describing the rotation between the coprecessing frame and a
specified inertial frame; (iii) the orbital phase in the coprecessing frame
$\phi_{\rm orb}$; and (iv) the precession of component spins $\bm{\chi}_1$,
$\bm{\chi}_2$ over time.

Modeling the BH remnant's properties is performed with the surrogate
surfinBH7dq2~\cite{Varma:2018aht}, which was also trained on the same set of NR
simulations.  This model takes in mass ratio $q$ and component spin vectors
$\bm{\chi}_1$, $\bm{\chi}_2$ at a given orbital frequency, and models the
remnant mass $m_f$, spin vector $\bm{\chi}_f$, and kick vector $\bm{v}_f$.

\subsection{Black-hole shapes}
\label{subsec:bh_shapes}

In our visualizations, we represent BH horizons with ellipsoids of revolution.
The axis of symmetry is along the instantaneous spin of the BH. The polar
(along the axis) and the equatorial (orthogonal to the axis) horizon radii are
set to
\begin{align}
    r_{\rm pol} = r_{+} \, , \qquad
    r_{\rm equi} = \sqrt{r_{+}^2 + a^2} \, ,
\end{align}
where $r_{+} = m + \sqrt{m^2 - a^2}$.  $r_{\rm pol}$ and $r_{\rm equi}$
correspond to the Kerr-Schild~\cite{Visser:2007fj, MTW} coordinate distances
from the BH center to the pole/equator of the horizon.  Note that numerical
simulations use a different coordinate system, meaning the BH shapes would be
different even for an isolated BH. However, this captures the azimuthal
symmetry and oblate nature seen in most coordinate systems.  

This approximation, however, neglects much of the interesting phenomenology of
event horizons (EHs) of BHs in binaries~\cite{baumgarteShapiroBook,
BohnMethods2016, BohnTorus2016}. Event horizons are defined globally, so the
locations of EHs cannot be determined without knowing the entire future
development of a spacetime.  Most NR simulations track the location of apparent
horizons (AHs)~\cite{baumgarteShapiroBook}, which \emph{can} be defined
locally.  Both EHs and AHs of orbiting BHs are deformed by the tidal field of
the other BH.  This distortion becomes very strong close to merger, where the
shape of the two event horizons do not resemble, even vaguely, that of
ellipsoids (see e.g.~\cite{SXSEventHorizonAnimations}).  Improving our
representation of EH shapes requires building surrogate models for the
morphology of the EH/AH, which is an interesting avenue for future work.

In addition, we assume the masses of the BHs are constant during the evolution.
While the masses in an NR simulation can change due to in-falling energy
through GWs, this is a very small effect (4PN (Post Newtonian) higher than
leading orbital energy loss~\cite{Poisson:1994yf,Alvi:2001mx,Poisson:2004})
that is safely
ignored in current waveform models including NRSur7dq2.

\subsection{Component black-hole spin evolution}
\label{subsec:bh_spins}

The two spins $\bm{\chi}_1,\bm{\chi}_2$ are modeled using NRSur7dq2. These are
known to agree well with NR simulations and are crucial for the accuracy of
that waveform model~\cite{Blackman:2017pcm}. Note, however, that the spins
modeled by NRSur7dq2 have had an additional smoothing filter applied to remove
short-timescale oscillations~\cite{Blackman:2017pcm}. This approximation
propagates to our visualizations. Similarly to the masses of the BHs, we assume
the spin magnitudes are constant during the evolution.  In-falling angular
momentum in the form of GWs can alter the spin magnitudes, but this is also a
very small effect (4PN higher than leading angular-momentum
loss~\cite{Alvi:2001mx,Poisson:2004}) that is ignored by current waveform
models including NRSur7dq2.

Spins are represented as arrows centered at the BH centers, that are
proportional to the Kerr parameter $\bm{a}$ of each BH. More specifically, the
length is set to $10a$, and the direction is along $\hat{\bm{a}}$. The
exaggeration of the magnitude is necessary to make the spin vectors clearly
visible during the evolution; more on this in the next section.

\subsection{Orbital angular momentum}
\label{subsec:orbital_ang_mom}

NRSur7dq2 only predicts the unit rotation quaternion $\hat{Q}(t)$  and not the
magnitude $L$. The (time dependent) direction of orbital plane is inferred from
$\hat{Q}(t)$ and is orthogonal to the z-axis of the coprecessing frame. For the
magnitude $L$, we implement the Newtonian expression
\begin{align}
  \label{Eq:Newtonian_L}
  L = M^2 \frac{q}{(1+q)^2} ~(M \omega_{\rm orb})^{-1/3},
\end{align}
where $\omega_{\rm orb}$ is the orbital frequency, as derived from the orbital
phase in the coprecessing frame modeled by \mbox{NRSur7dq2},
\begin{align}
  \label{eq:omega_orb}
  \omega_{\rm orb} = \frac{d \phi_{\rm orb}}{dt} \,.
\end{align}

In our visualizations, the angular momentum is indicated by an arrow at the
origin. Its magnitude is rescaled to $12L$.  This factor is arbitrary and it is
chosen to make the arrow clearly visible.

Note that it is not appropriate to compare an arrow for orbital angular
momentum $L\propto M^2$ to those representing the Kerr parameters
${a}_1,{a}_2\propto M$ because they have different dimensions. The choice of
representing $a$, rather than the $S \propto M^2$ was made to allow all arrows
to be clearly visible throughout the inspiral for generic locations in the
parameter space (i.e.\ different mass ratios). However, we provide an option to
represent $S$ for the spin arrows (cf.  Sec.~\ref{sec:python_implementation}),
in which case the arrow magnitudes are set to $12S$. This makes the arrow on
the smaller BH barely visible in some cases, but allows direct comparison of
the spin arrows to the orbital angular momentum arrow.  This could be
informative for gaining intuition about peculiar spin phenomena like
transitional precession~\cite{Apostolatos1994,Zhao:2017tro}, spin orbit
resonances \cite{2004PhRvD..70l4020S}, large nutations
\cite{2015PhRvL.114n1101L,2018arXiv181105979G} and precessional instabilities
\cite{2015PhRvL.115n1102G}. This phenomenology is currently beyond the scope of
the surrogate we used, but is being actively researched with NR simulations
\cite{2018PhRvD..98h3014A, Inprep-Ossokine-et-al:2019} and lies within the
realm of future hybridized surrogate models (see e.g.~\cite{Varma:2018mmi}).

\subsection{Component black-hole trajectories}
\label{subsec:trajectories}

The gauge symmetry of GR is broken in an NR simulation, since one necessarily
has to specify a set of coordinates to represent the solution on a computer.
The BH trajectories extracted from numerical simulations are, therefore,
inherently gauge dependent.

In the construction of NRSur7dq2~\cite{Blackman:2017pcm} quantities  like
$\hat{Q}(t)$ and $\phi_{\rm orb}$ are obtained from the GWs extrapolated to future
null infinity, not from numerical simulations' BH coordinates.

In our visualizations, we reconstruct the trajectories of the BHs using the
dynamics predicted by NRSur7dq2 and some PN arguments.  In particular, one
needs the separation between the BHs as a function of the orbital frequency,
$r_{12}(\omega_{\rm orb})$, with the orbital frequency defined as in
Eq.~\eqref{eq:omega_orb}. The separation $r_{12}(\omega_{\rm orb})$ is modeled
using the 3.5PN expressions reported in Eq.~(4.3) of Ref.~\cite{Bohe:2012mr},
along with the 2PN spin-spin term from Eq.~(4.13) of Ref.~\cite{Kidder:1995zr}.

Let us write the coprecessing frame coordinates as $(x', y', z')$.  The
trajectories in the coprecessing frame, where the orbital plane is orthogonal
to the $z'-$axis, are given by
\begin{align}
\begin{cases}
x'_{1} = r_1 \cos\phi_{\rm orb} \\
y'_{1} = r_1 \sin\phi_{\rm orb} \\
z'_{1} = 0
\end{cases}
\begin{cases}
x'_{2} = - r_2 \cos\phi_{\rm orb} \\
y'_{2} = - r_2 \sin\phi_{\rm orb} \\
z'_{2} = 0
\end{cases}
\end{align}
where $r_{1}$ ($r_{2}$) indicates the coordinate separation from the origin to
the primary (secondary) BH center. We use the Newtonian relations
\begin{equation}
    r_{1} = \frac{m_{2}}{M} ~r_{12},
    \qquad
     r_{2} = \frac{m_{1}}{M} ~r_{12},
\end{equation}
to enforce the Newtonian center-of-mass of the binary to be at the origin. This
ignores the fact that true center of mass during inspiral and merger oscillates
about the origin due to linear momentum carried away in GW. However, this
correction would be too small to be noticeable on the scale of our
visualizations (see e.g. Fig.~2 of \cite{Gerosa:2018qay}).

\begin{figure}
\begin{center}
\includegraphics[scale=0.8]{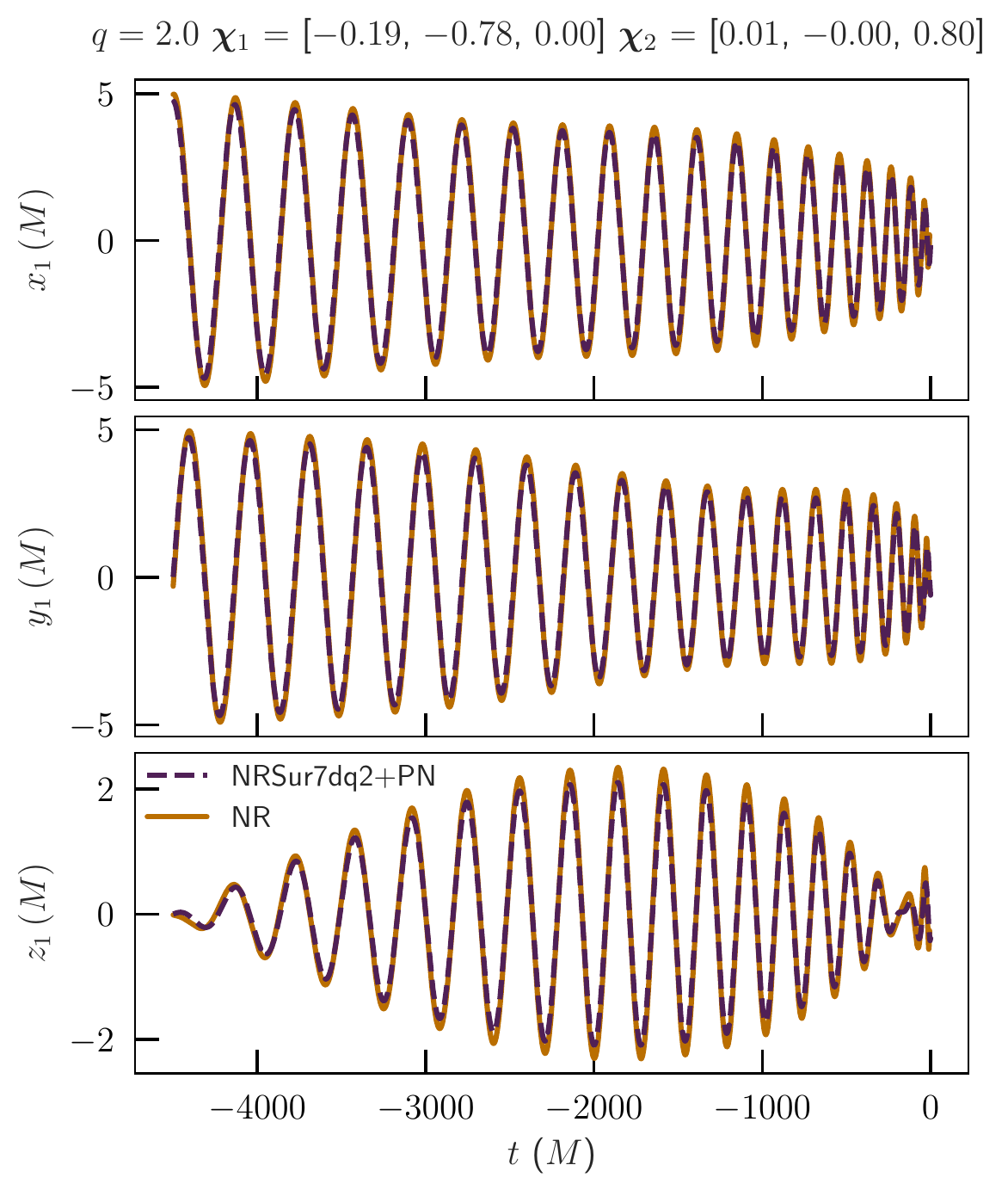}
\end{center}
\vspace{-1em}
\caption{Comparison of the coordinate trajectories of the heavier BH for a
precessing binary BH, between NR, and our approximation using NRSur7dq2 and PN.
$t=0$ corresponds to the peak of the waveform amplitude. The mass ratio, and
spins at $t=-4500M$ are shown at the top of the plot. 
} 
\label{fig:trajectories}
\end{figure}

Given the trajectories in the coprecessing frame, the trajectories in the
inertial frame are obtained by a quaternion transformation with the
time-dependent rotation (unit) quaternions $\hat{Q}(t)$ (for a brief
introduction to quaternions in this context, see e.g.\ App.\ A
of~\cite{Boyle:2013a}).  Treating the Euclidean positions as purely
imaginary quaternions, the transformation is
\begin{align}
  \bm{x}_{i} = \hat{Q}(t)~ \bm{x}'_{i} ~\hat{Q}^{-1}(t)\,.
\end{align}

Figure~\ref{fig:trajectories} compares the trajectories predicted by our method
to the gauge-dependent ones extracted from an NR simulation. Our approximate
trajectories turn out to be remarkably close to the NR trajectories.  The
dominant deviations are due to the PN formulae being in harmonic gauge, whereas
the NR simulations use the damped harmonic gauge~\cite{Szilagyi:2009qz}.

\subsection{Gravitational waves}
\label{subsec:gw_methods}

NR simulations predict the entire spacetime metric of a binary BH evolution.
However, the full metric is usually discarded because most applications
(notably GW observations) only require the gravitational waves as seen by an
observer far away.

Indeed, splitting the metric into GWs and a non-oscillatory part can only be
well defined in the \emph{wave zone}, which is at distances $r$ much larger
than the gravitational wavelength $\lambda$.  Let us suppose we are in a
spacetime that is approximately Minkowski space, with a metric perturbation
$h_{ab}$, in the transverse-tracefree (TT) gauge~\cite{Flanagan:2005yc}.  We
define a spherical polar coordinate system $(t,r,\theta,\phi)$ with the binary
center-of-mass at the origin.  The $z$ axis ($\theta=0$) of this coordinate
system is parallel to $\bm{L}$ at some reference time/frequency. The $x$ axis
lies along the line of separation from the lighter BH to the heavier BH at this
time/frequency, and the $y$ axis completes the triad.

The spherically outgoing gravitational wave is typically converted into a
spin-weight $-2$ complex scalar by contracting $h \equiv
h_{ab}\bar{m}^{a}\bar{m}^{b}$, where $m^{a} = (\hat{e}_{\theta}^{a}+ i
\hat{e}_{\phi}^{a})/\sqrt{2}$ is an element of a complex null dyad~\cite{MTW} along
with its conjugate $\bar{m}^{a}$; and where $\hat{e}_{\theta}^{a}, \hat{e}_{\phi}^{a}$ are
the standard unit vectors in the $\theta$ and $\phi$ directions, respectively.
The gravitational-wave strain $h$ is then decomposed as
\begin{equation}
    h(t, r,\theta,\phi) = \sum_{\ell=2}^{\infty} \,\sum_{m=-\ell}^{\ell}
    {}_{-2}Y_{\ell m}(\theta,\,\phi)
    \, h_{\ell m}(t, r)
    \,,
\label{Eq:mode_sum}
\end{equation}
where $_{-2}Y_{\ell m}$ are the $s\!=\!-2$ spin-weighted spherical
harmonics~\cite{Goldberg1967}.  The functions $h_{\ell m}$ are referred to as
the \emph{modes} of the GWs.

From the structure of the flat-space d'Alembertian operator, we can see that at
large distances, $h$ is dominated by a piece decaying as $\sim 1/r$ along lines
of constant retarded time $t_{ret} \equiv t-r$ \cite{thorne80}.  This motivates
how waves are extracted from NR.  First, $(r h_{\ell m})$ is evaluated on
spheres of various radii in the computational domain.  This is then
extrapolated to future null infinity, defining
\begin{gather}
    (rh_{\ell m})^{\infty}(t) \equiv \lim_{r \to \infty} r~ h_{\ell m}(t-r, r).
\end{gather}
NRSur7dq2 only models these extrapolated GW modes, $(rh_{\ell m})^{\infty}$.

One can evaluate the GWs at any particular orientation in the source frame at
$r\to \infty$ by applying Eq.~(\ref{Eq:mode_sum}) to $(rh_{\ell
m})^{\infty}(t)$.  This is used to generate the waveform time series in the
bottom subplots of our animations (cf.\ Fig.~\ref{fig:visualization_example}),
where we show the plus $h_{+} = \Re(h)$ and cross $h_{\times} = - \Im(h)$
polarizations.  We use all the spin-weighted spherical harmonic modes provided
by NRSur7dq2, i.e.\ $2 \leq \ell \leq 4$ and $|m|\leq \ell$.

Since the full metric is not available in the bulk, we approximate it from
$(rh)^{\infty}$.  When showing GWs on the bottom plane of our visualizations
(cf.~Fig.~\ref{fig:visualization_example}), we approximate the strain as
\begin{gather}
  h(t,r,\theta,\phi) \approx \frac{(rh)^{\infty}(t_{ret},\theta,\phi)}{r}
  \,.
\label{Eq:bulk_waveform}
\end{gather}
This neglects curved-background effects such as tails, and higher order $1/r$
corrections, so this approximation is only valid at large $r$.  More work would
be needed to recover the higher powers of $1/r$, but it is technically possible
(see Eq.~(2.53a) of~\cite{thorne80}). The default position of the bottom-plane
is quite close to the binary; moving it farther out improves this
approximation.

\subsection{Post merger phase}

In NR simulations, a common apparent horizon typically forms at a
retarded time close to
the peak of the waveform $\mathcal{A}^2=\sum_{\ell,m} | h_{\ell m}|^2$.  This
is taken to be the definition of the time of merger.  We therefore shift the
time variable $t$ such that $t\!=\!0$ corresponds to $\max_t\mathcal{A}$.  At
$t\geq0$, the two component BHs are replaced by a single remnant.  The final
mass, spin, and kick of the remnant are predicted using
surfinBH7dq2~\cite{Varma:2018aht}.

Mass and spins of the remnant are used to draw a horizon ellipsoid and spin
arrow as specified in Sec.~\ref{subsec:bh_shapes} and
Sec.~\ref{subsec:bh_spins}.  The remnant BH horizon is expected to be highly
distorted at the common horizon formation time.  We ignore this effect and
simply represent the remnant BH by an ellipsoid of constant shape from $t\!=\!0$
onwards.

During a BH inspiral and merger, linear momentum emitted in GWs causes motion
of the binary's center of mass (cf.\ e.g.\ Ref.~\cite{Gerosa:2018qay} and
references therein). In practice, however, linear momentum flux is negligible
at early times and the ``kick'' is only accumulated over the last few cycles
before merger.  Here we make the additional simplification of neglecting this
effect, and assume that the remnant is formed at the origin and receives all of
its kick velocity instantaneously. However, as mentioned before, this
correction would be at a scale that is not noticeable in our visualizations
(cf. Fig.~2 of \cite{Gerosa:2018qay}).

\subsection{Time steps and displayed text}

To better highlight different phases of the evolution, we use a non-uniform
time step.  The time step between frames  at $t\lesssim75 M$ is chosen to
obtain $30$ frames for each orbit. The animation, therefore, is artificially
slowed down close to merger, so that the entire dynamics is easier to observe.
After the ringdown stage, the animation is sped up to better illustrate the
final kick. The current time is displayed in the figure text, as well as
indicated by the blue vertical slider in the bottom waveform subplot (cf.
Fig.~\ref{fig:visualization_example}).

The figure text at the top-left of the main visualization panel shows the
parameters of the binary (remnant). At times $t<0$, these are the mass ratio
and instantaneous spin components. Mass, spin and kick of the remnant BH are
shown after merger.

\begin{figure*}[bth]
\begin{center}
\makebox[\linewidth][c]{%
\includegraphics[scale=0.6]{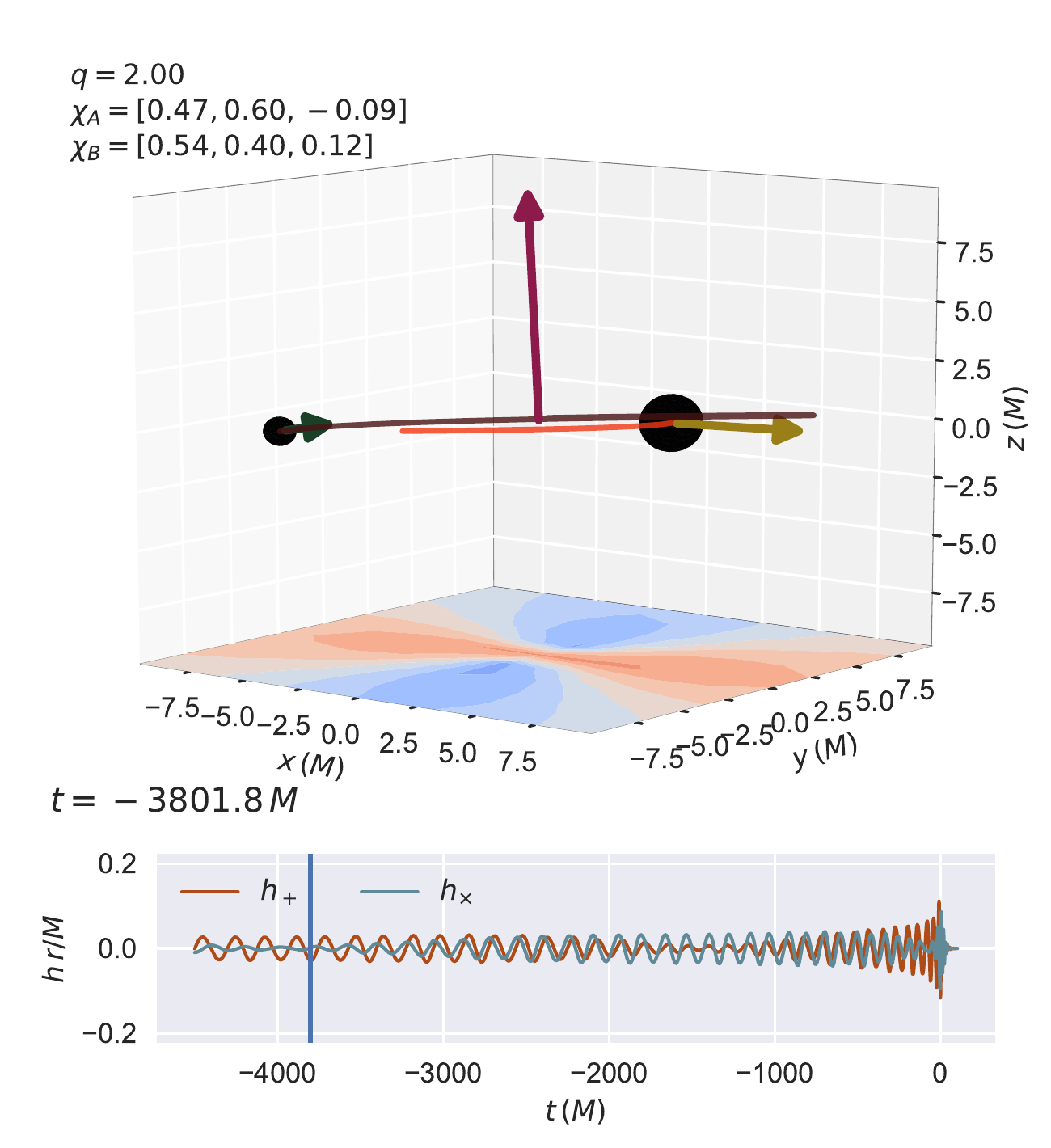}
\includegraphics[scale=0.6]{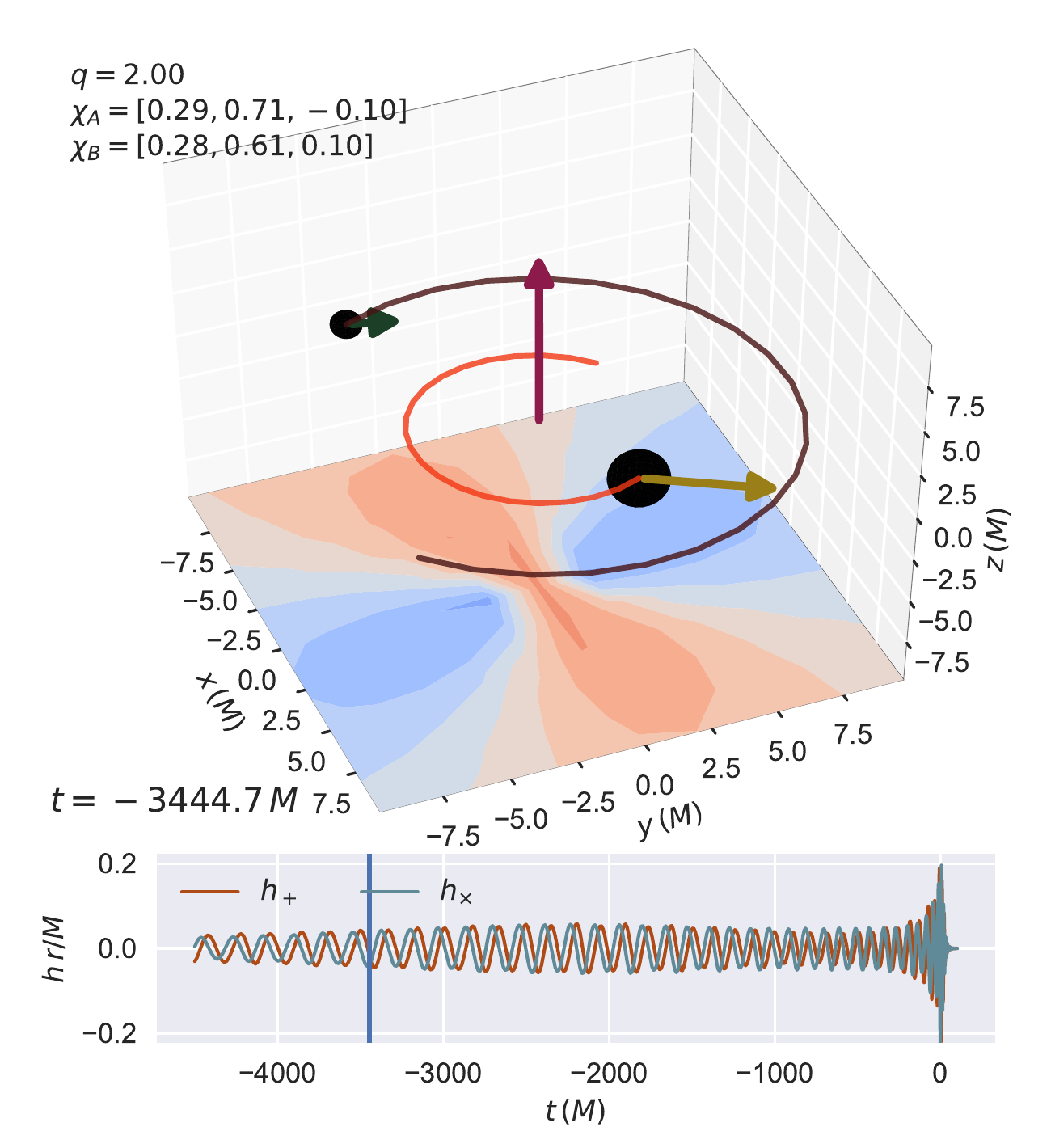}
}
\includegraphics[scale=0.6]{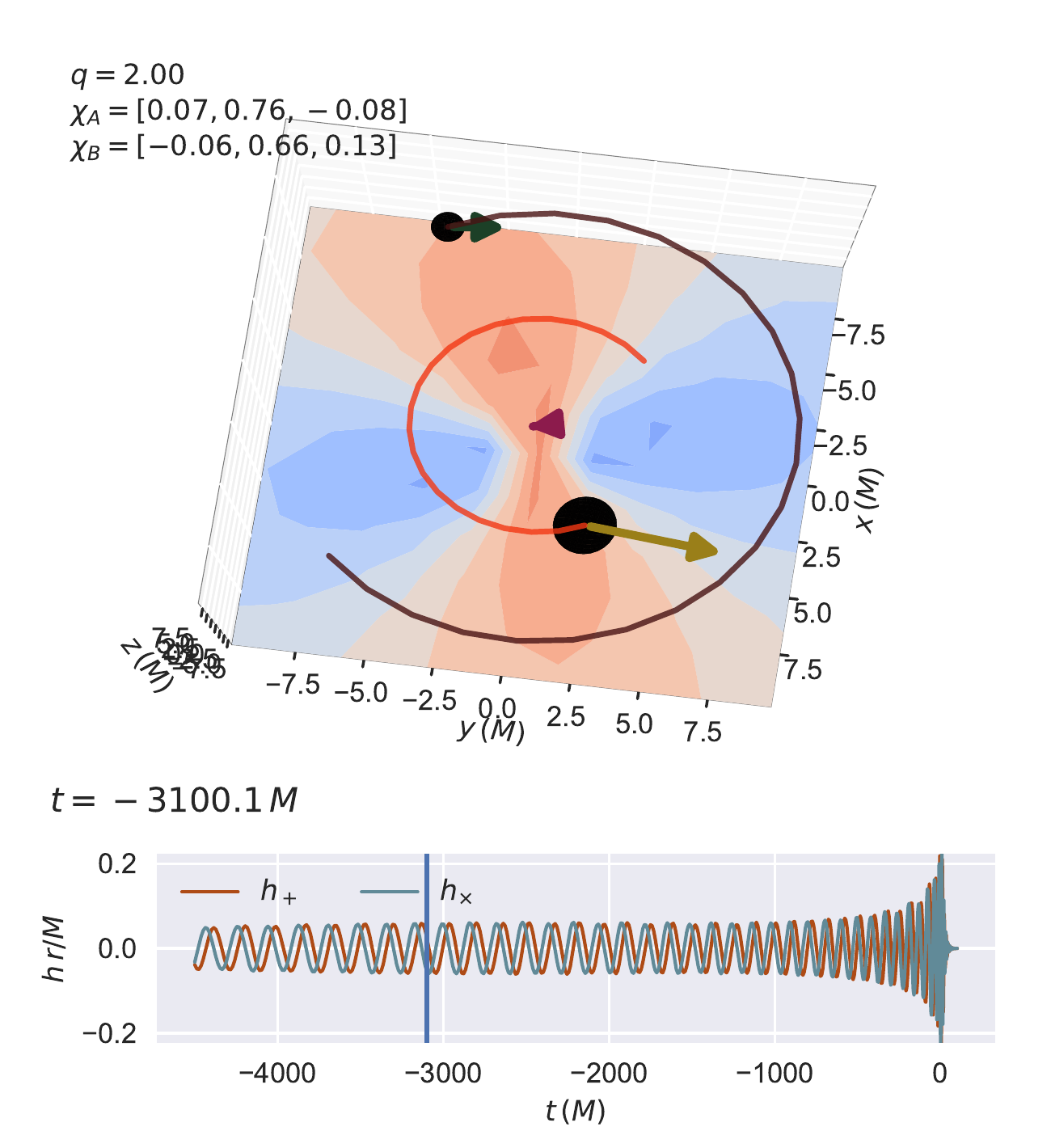}
\end{center}
\vspace{-1em}
\caption{Visualization of a precessing binary black hole system where we also
vary the camera viewing angle during the inspiral. Notice how the waveform
structure in the bottom subplots changes based on whether the viewing angle is
edge-on (top-left), intermediate (top-right), or face-on (bottom).  This
animation is available at
\href{https://vijayvarma392.github.io/binaryBHexp/\#prec_bbh_rotating_camera}{vijayvarma392.github.io/binaryBHexp/\#prec\_bbh\_rotating\_camera}.
}
\label{fig:waveform_projections}
\end{figure*}
\section{Explorations}
\label{sec:explorations}

We now provide additional examples that demonstrate the power and utility of
our visualizations.

\subsection{Waveform projection}

Figure \ref{fig:waveform_projections} shows a visualization of a precessing
binary BH, when we also vary the camera viewing angle during the evolution. The
polarization content and the morphology of the waveform therefore strongly
depend on the direction of the line of sight, which can be understood as
follows. From Eq.~\eqref{Eq:mode_sum}, the observer viewing angles $(\theta,
\phi)$ affect the relative weights with which the waveform modes $h_{\ell m}$
are combined into the strain $h$.  Note that the standard quadrupole formula
for GW emission only contains the dominant $\ell\!=\!|m|\!=\!2$ modes, while
here we use all modes with $\ell\leq4$. 

The GW amplitude is strongest along the direction of $\bm{L}$. This is evident
from the bottom panel of Fig.~\ref{fig:waveform_projections}, where the
direction of $\bm{L}$ aligns with the observer's viewing angle (i.e., the
binary is \emph{face-on}). On the other hand (top-left panel of
Fig.~\ref{fig:waveform_projections})  the GW amplitude is at its least when the
observer viewing angle is orthogonal to $\bm{L}$ (\emph{edge-on}).  The
contribution of higher harmonics $\ell>2$ to Eq.~(\ref{Eq:mode_sum}) also
depends on observer viewing angle.  For face-on binaries, the GWs are strongly
dominated by the quadrupolar modes. Going from face-on to edge-on, the
contribution of the quadrupolar modes decreases and that of the nonquadrupolar
modes increases.

One can also infer the polarization content of the GWs from the waveform panel.
If there is a $\pm 90^{\circ}$ phase shift between $h_{+} = \Re(h)$ and
$h_{\times} = -\Im(h)$, the GWs are circularly polarized.  The bottom panel
of Fig.~\ref{fig:waveform_projections}, which is mostly face-on, shows almost
perfect circular polarization, deviating due to precession of the orbital
plane.  For comparison, when $h_{+}$ and $h_{\times}$ are proportional with a
real constant of proportionality, the GW has a linear polarization (this
includes the simpler case where one of the two polarizations vanishes).  The
top-left panel of Fig.~\ref{fig:waveform_projections}, where the system is
(almost) edge-on, exhibits (almost) linear polarization at many times
throughout the inspiral.  Again the deviations are due to precession of the
orbital plane.  The modulation is more noticeable for nearly edge-on precessing
systems, since one of the polarizations can temporarily vanish as the system
precesses through perfectly edge-on configurations.

\newcommand{\huplot}[1]{\raisebox{-.5\height}{\includegraphics[scale=0.16]{hangup_t#1}}}
\newcommand{\rothulabel}[1]{\raisebox{-.5\height}{\rotatebox{90}{{\footnotesize $ t\sim#1\,M$}}}}
\begin{figure*}[p]
\makebox[\linewidth][c]{%
  \begin{tabular}{cc}
    \raisebox{-.5\height}{\rotatebox{90}{{\footnotesize $\omega_{\text{orb}}=0.018~\text{rad}/M$}}}
    &\huplot{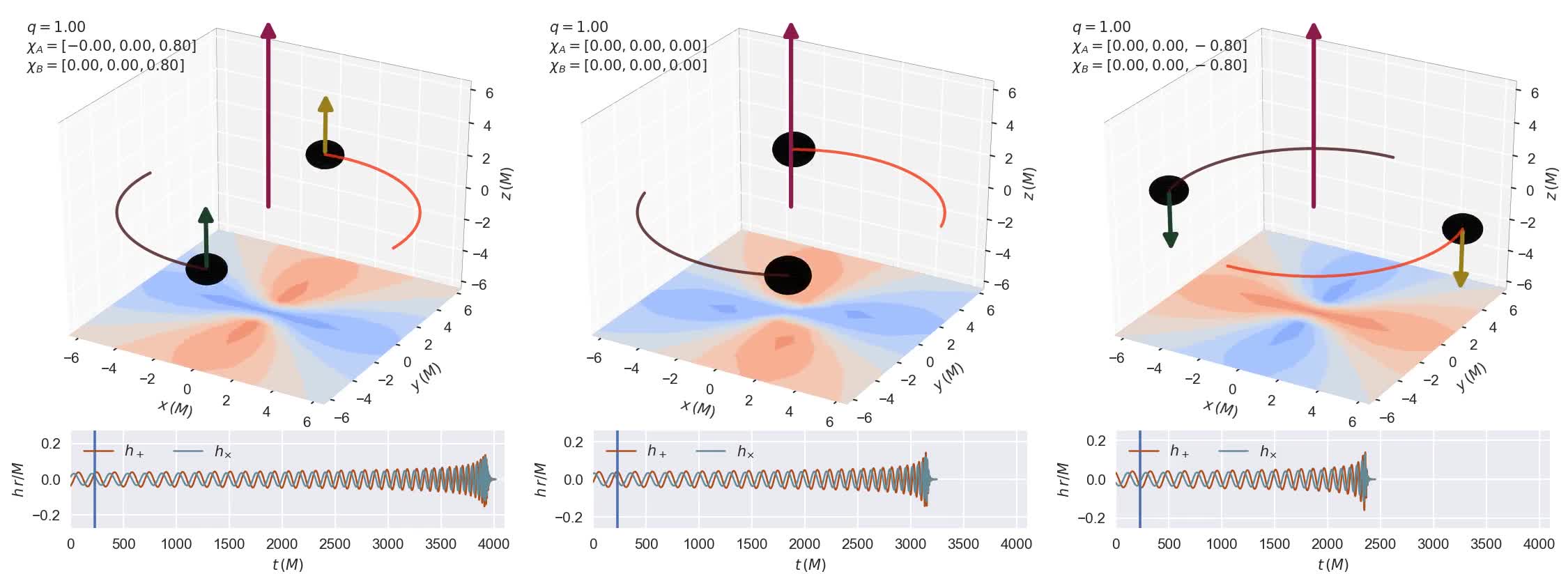}\\
\rothulabel{2300}&\huplot{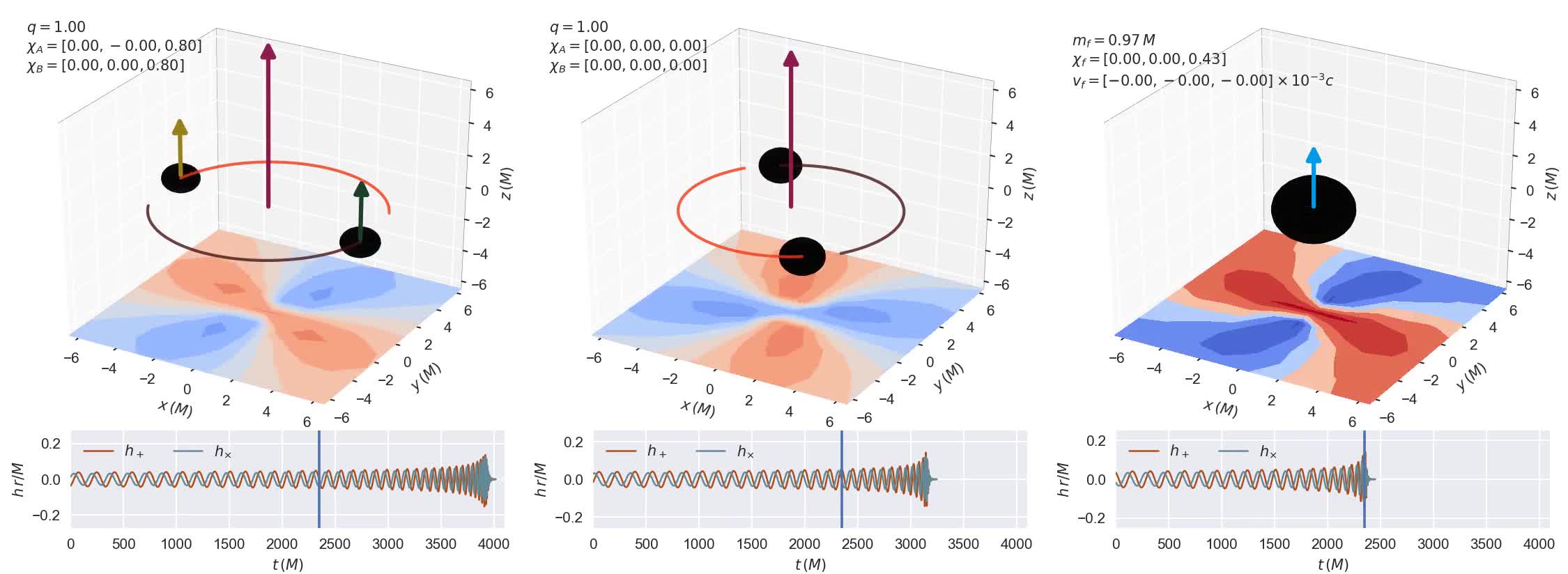}\\
\rothulabel{3150}&\huplot{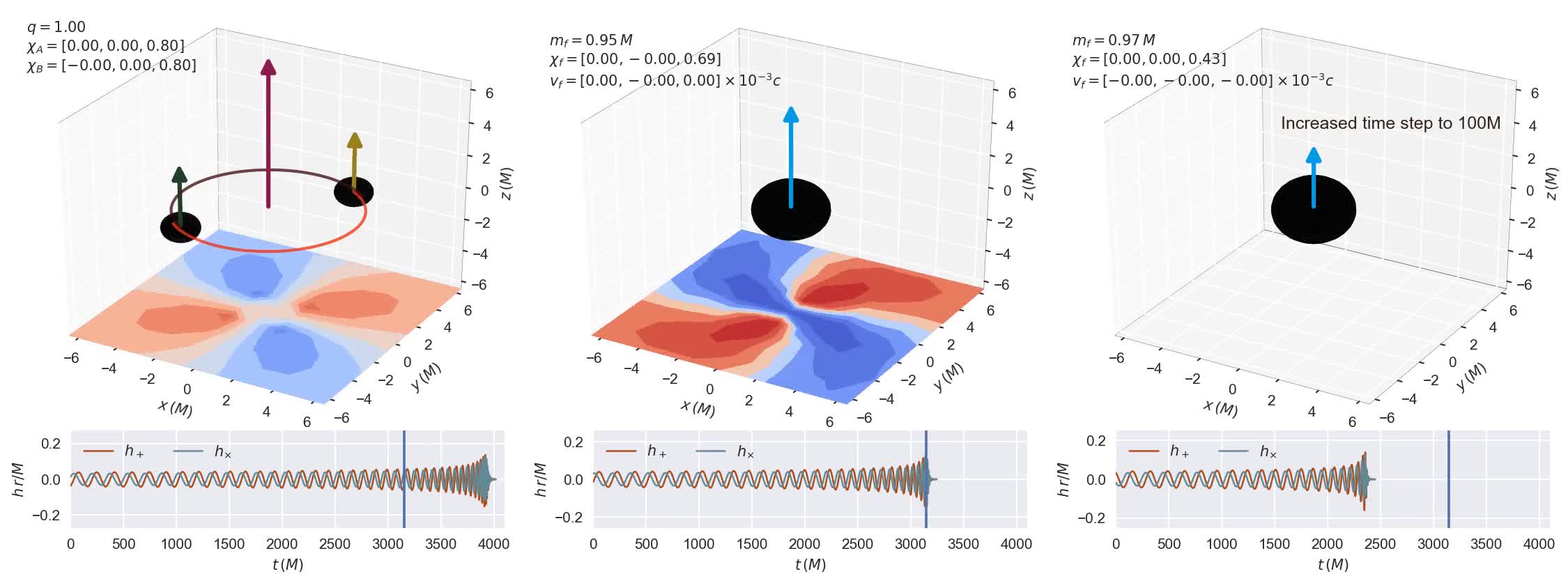}\\
\rothulabel{3950}&\huplot{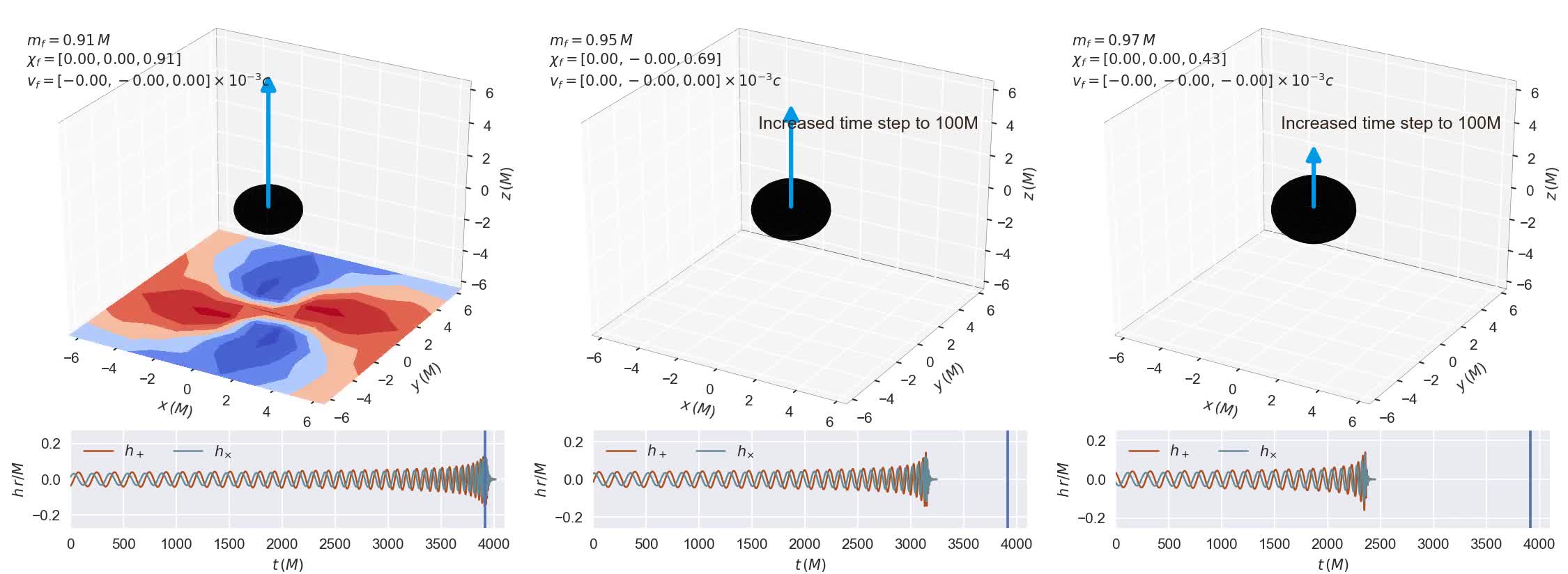}
  \end{tabular}
}
\caption{Visualization of the orbital hang-up effect. We show three
nonprecessing systems with equal masses, and equal spins. In the left (right)
column, both spins are aligned (anti-aligned) with $\bm{L}$, with magnitude
0.8.  The middle column shows a nonspinning binary. All three systems start at
an orbital frequency of $0.018~\text{rad}/M$. Due to orbital hang-up
effect, the length of the waveform is longer (shorter) for the aligned case
compared to the nonspinning case (see the bottom subplots showing the
waveform). Time flows downwards (labeled at the left), and each row corresponds
to a fixed time since the start of the animation. This animation is available
at
\href{https://vijayvarma392.github.io/binaryBHexp/\#hangup}{vijayvarma392.github.io/binaryBHexp/\#hangup}.
}
\label{fig:hangup}
\end{figure*}
\subsection{Orbital hang-up effect}
Apart from precession, the BH spins have other important effects on the
evolution of binaries. One such effect is the so called \emph{orbital hang-up}
effect~\cite{Damour:2001tu,Campanelli2006c,Scheel:2014ina} which delays or
prompts the merger of the BHs based on the sign of the BH spin component along
the orbital angular momentum, $\bm{S} \cdot \bm{L}$, where $\bm{S}$ is one of
$\bm{S}_1$ or $\bm{S}_2$.  This spin-orbit coupling is a $1.5$ PN effect that
effectively acts as an additional repulsion (attraction) when the sign of
$\bm{S} \cdot \bm{L}$ is positive (negative). This means that binaries that
have spins that are aligned (anti-aligned) with $\bm{L}$ will merge slower
(faster) than nonspinning binaries, when starting from the same orbital
frequency.  This is analogous to the location of the innermost stable circular
orbits of Kerr BHs, which is at a smaller (larger) radius for
co-(counter-)rotating particles.

This is demonstrated in Fig.~\ref{fig:hangup}, which shows an aligned,
nonspinning and an anti-aligned binary, starting at the same orbital frequency.
Unlike the rest of the animations discussed in this paper, here we use a
constant time step between the frames of the movie (rather than a fixed 30
frames per orbit), and set $t=0$ at the start of the waveform (rather than at
the peak).  Due to the orbital hang-up effect, the anti-aligned binary merges
first, followed by the nonspinning system, and finally the aligned system. In
addition, the aligned (anti-aligned) binary radiates more (less) energy due to
its prolonged (shortened) evolution, and the final mass is therefore smaller
(larger) than the nonspinning case.  The interaction between spin and orbital
angular momentum also determines the remnant spin in a non-trivial way: the
aligned (anti-aligned) case results in the largest (smallest) remnant spin
magnitude.

The orbital hang-up effect can also be explained heuristically using the cosmic
censorship conjecture. For the aligned-spin binary in Fig.~\ref{fig:hangup},
the initial magnitude of total angular momentum is given by $J=L + m_1^2~\chi_1
+ m_2^2~\chi_2$. Using $L$ from Eq.~(\ref{Eq:Newtonian_L}) with $\omega_{\rm
orb}=0.018~\text{rad}/M$, we get $J\sim1.35 M^2$.  This is larger than the
maximum allowed spin angular momentum for a Kerr BH, $M^2$. On the other hand,
for the anti-aligned case we have  $J=L -m_1^2~\chi_1 -m_2^2~\chi_2 \sim 0.55
M^2$, which is well within the limit. So, the aligned binary must radiate at
least $0.35 M^2$ of its total angular momentum in the form of GWs before it can
merge, in order to not violate cosmic censorship. The anti-aligned case can
therefore merge faster.

\begin{figure*}[p]
\includegraphics[width=0.5\textwidth]{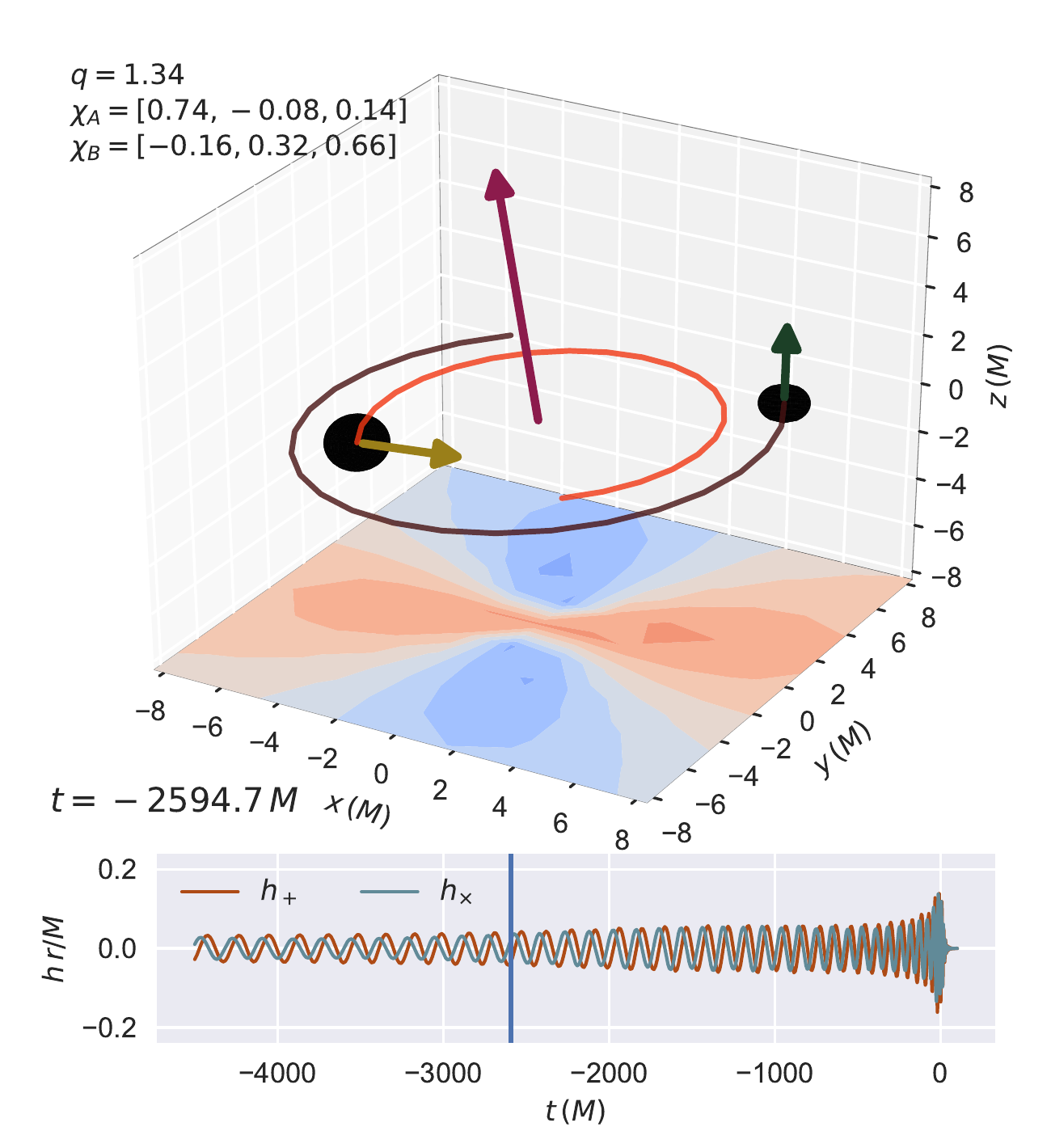}
\includegraphics[width=0.5\textwidth]{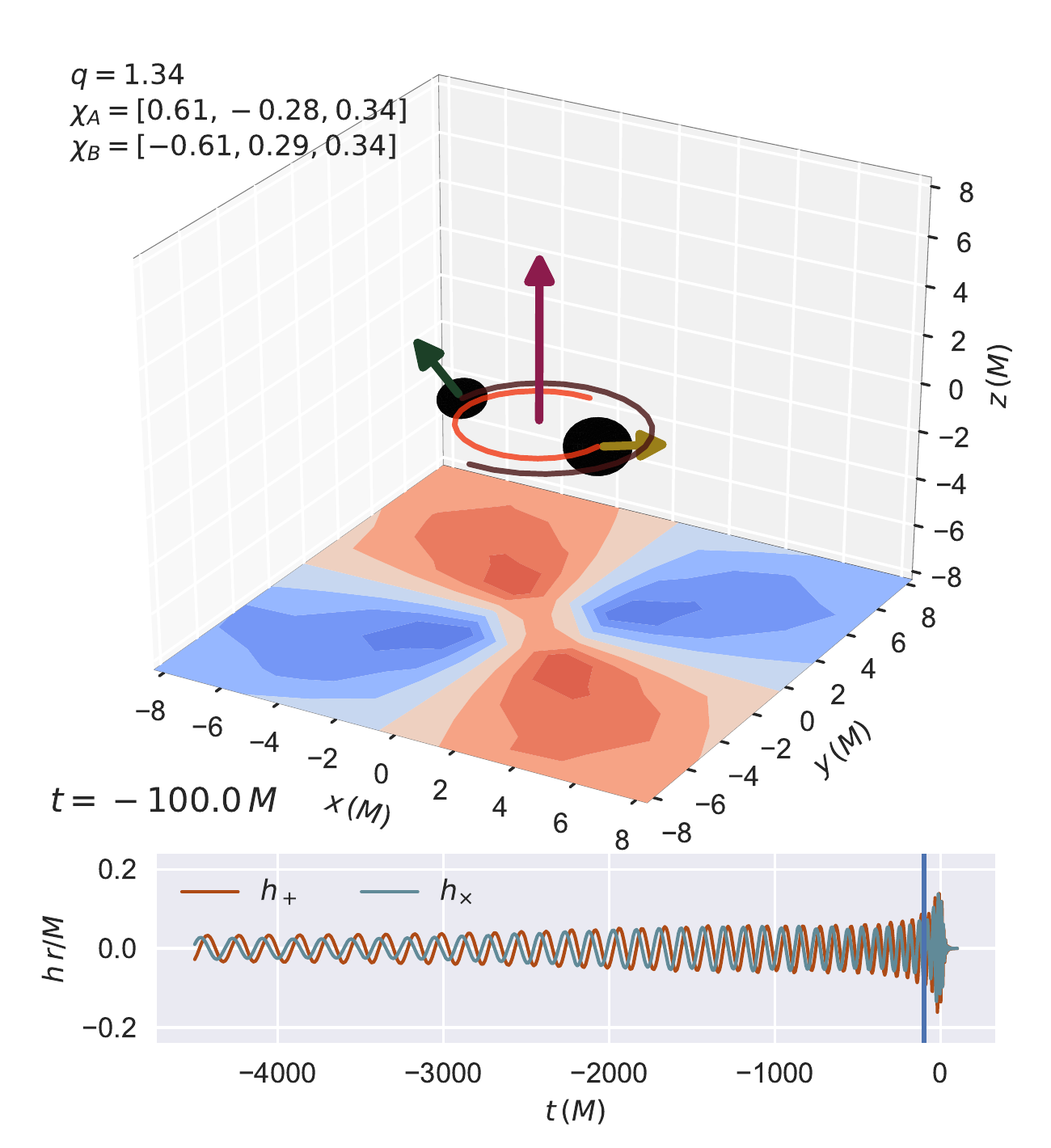}  \\
\includegraphics[width=0.5\textwidth]{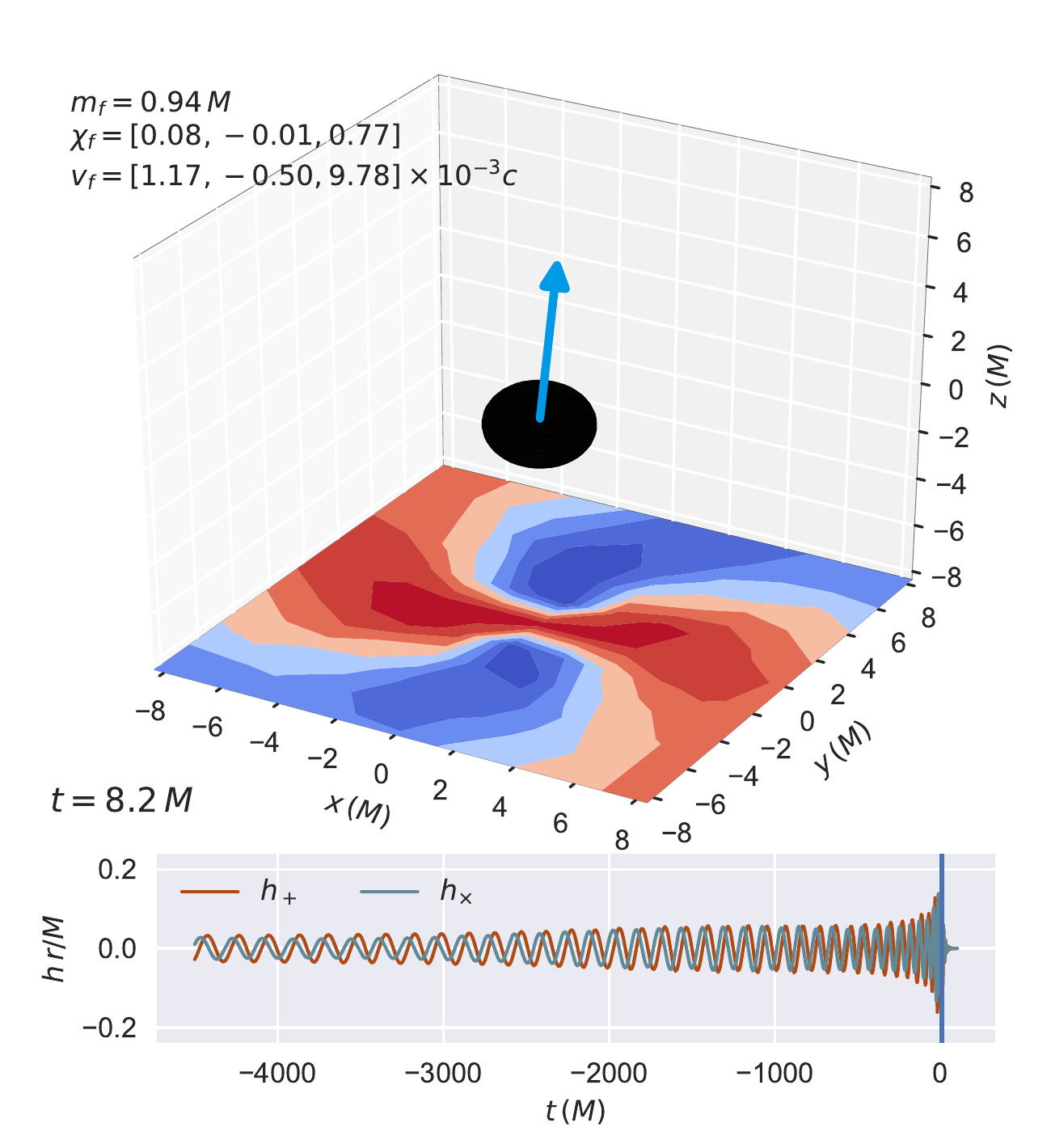}
\includegraphics[width=0.5\textwidth]{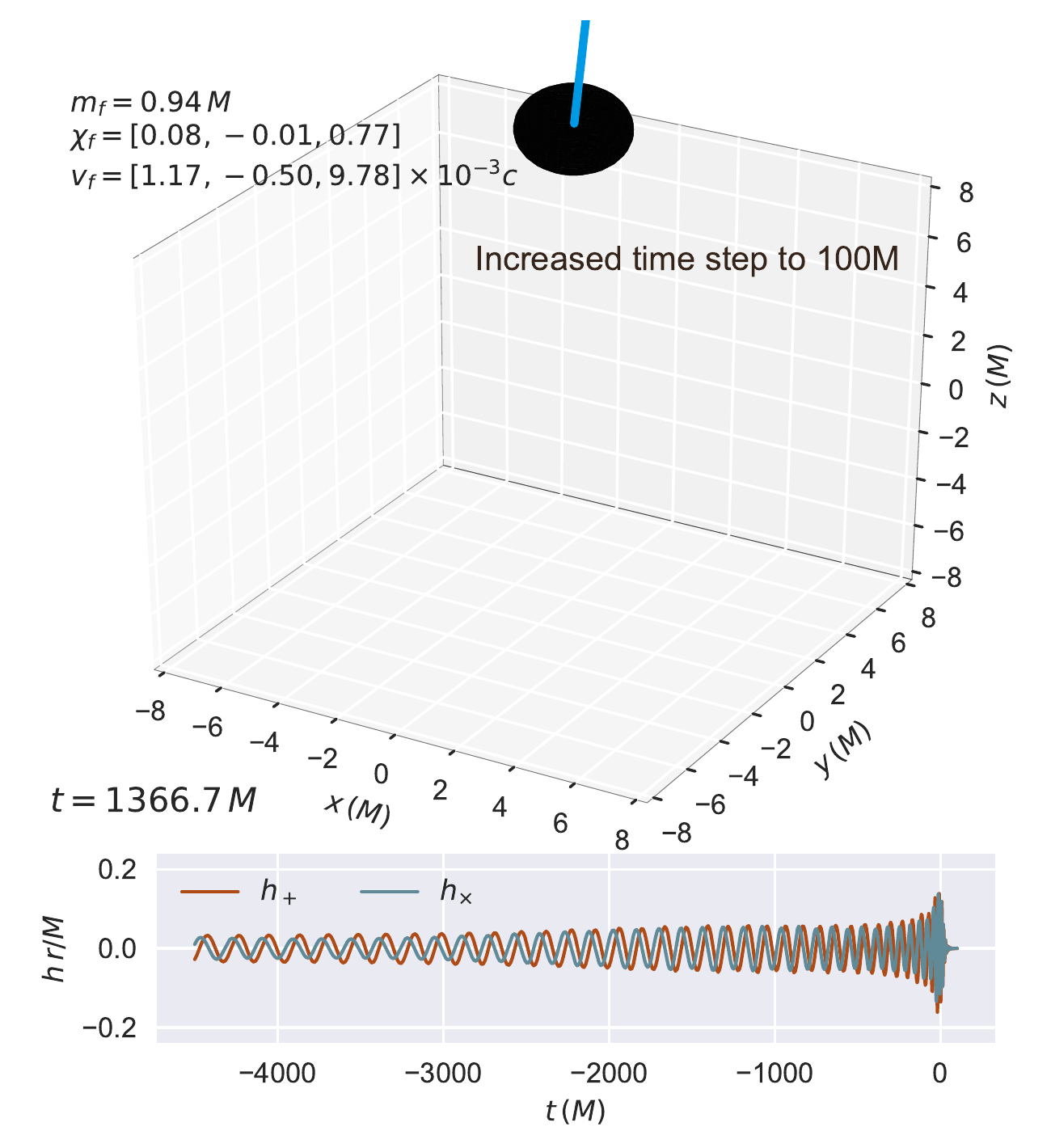}
\caption{Evolution of a super-kick configuration. Time flows from left to right
and from top to bottom, as shown at the bottom left of each panel.  The
top-left panel shows a snapshot taken in the early inspiral. In the top-right
panel, the two BHs are about to merge and the spins are are seen to be in a
super-kick configuration.  The bottom-left snapshot is taken at the time at
which the peak of the waveform hits the bottom plane where the GW pattern is
shown. After merger (bottom-right panel), the final BH is imparted a kick of
$\sim 3000$ km/s (note that we speed up the animation after the ringdown by
increasing the time steps to $100M$).  This animation is available at
\href{https://vijayvarma392.github.io/binaryBHexp/\#super_kick}{vijayvarma392.github.io/binaryBHexp/\#super\_kick}.
}
\label{fig:super_kick}
\end{figure*}
\subsection{Super-kick}

Next, we consider a binary BH in the so-called \emph{super-kick} configuration.
Anisotropic emission of GWs causes a net flux of linear momentum, which imparts
a kick to the remnant BH. Some degree of asymmetry is necessary for a nonzero
kick~\cite{Boyle2007a}. For instance the kick vanishes by symmetry during the
merger of an equal-mass, nonspinning binary BH system.  Strongly precessing
binary BHs have been found to generate the highest kicks~\cite{Campanelli2007a,
Gonzalez2007b, Lousto:2011kp}. Some of these systems have kicks large enough to
escape from even the most massive galaxies in the Universe~\cite{Merritt2004,
Gerosa:2014gja}.

In particular, a vary large kick (up to $\sim 3000$ km/s) is imparted to BHs
merging with spins lying in the orbital plane and anti-parallel to each other.
These are the so-called \emph{super-kicks} first discovered in
2007~\cite{Campanelli2007a, Gonzalez2007b}, by means of NR simulations. The
largest kicks observed in numerical simulations to date are the so-called
\emph{hangup-kicks}~\cite{Lousto:2011kp}, where the spins have non-zero
components perpendicular to the orbital plane, but the in plane spins are
anti-parallel. We will refer to all configurations where the spins near merger
are coplanar, and their orbital plane projections are anti-parallel, as
super-kick configurations. Crucially, large kicks are only found if the spins
are in these fine-tuned configurations ``near merger.''

For this reason, generating visualizations of BH super-kicks from simulations
can be challenging. The spins are usually specified at the start of the
simulations and several attempts are necessary to find the specific initial
conditions that will result in co-planar spins near merger.  With our tool,
on the other hand, one can specify the spins at any time/frequency, including
close to merger.  Generating a visualization of a system in a super-kick
configuration is as easy as any other location in parameter space. This is
shown in Fig.~\ref{fig:super_kick}. The remnant reaches a final velocity  of
$\sim10^{-2} c$ ($\sim 3000 km/s$), in agreement with~\cite{Campanelli2007a,
Gonzalez2007b, Lousto:2011kp}.

\newcommand{\skplot}[2]{\raisebox{-.5\height}{\includegraphics[scale=0.2]{sine_plots/sinus#1_#2}}}
\newcommand{\rottlabel}[1]{\raisebox{-.5\height}{\rotatebox{90}{{\footnotesize $t=#1M$}}}}
\begin{figure*}[thb]
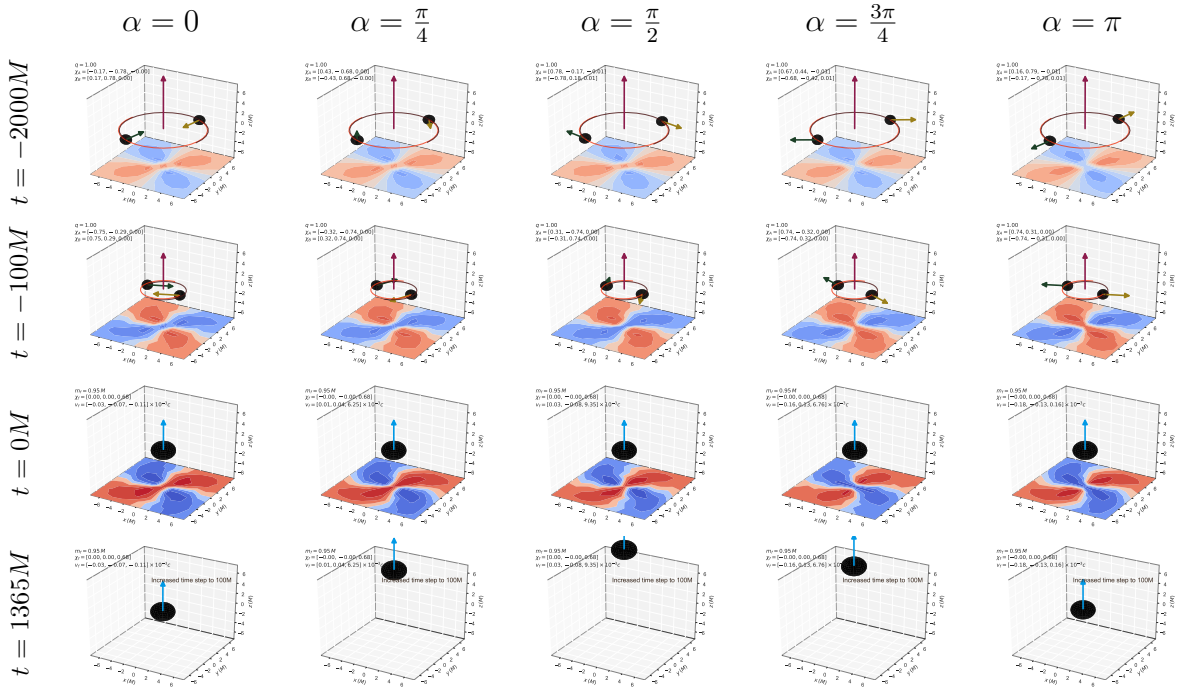

  \begin{tabular}{cccccc}
    &
    $\alpha=0$&
    $\alpha=\frac{\pi}{4}$&
    $\alpha=\frac{\pi}{2}$&
    $\alpha=\frac{3\pi}{4}$&
    $\alpha=\pi$\\
    \noalign{\smallskip}

\rottlabel{-2000}&
\skplot{0}{m2000}&
\skplot{1}{m2000}&
\skplot{2}{m2000}&
\skplot{3}{m2000}&
\skplot{4}{m2000}
\\

\rottlabel{-100}&
\skplot{0}{m100}&
\skplot{1}{m100}&
\skplot{2}{m100}&
\skplot{3}{m100}&
\skplot{4}{m100}
\\

\rottlabel{0}&
\skplot{0}{0}&
\skplot{1}{0}&
\skplot{2}{0}&
\skplot{3}{0}&
\skplot{4}{0}
\\

\rottlabel{1365}&
\skplot{0}{1365}&
\skplot{1}{1365}&
\skplot{2}{1365}&
\skplot{3}{1365}&
\skplot{4}{1365}
  \end{tabular}
\caption{%
Sinusoidal dependence of the kick magnitude on the angle between spins close to
merger. Five different cases are shown (left to right), with equal masses and
equal spins. Both spins are confined to the orbital plane, and are
anti-parallel to each other, but with a different angle in the plane $\alpha$
(labeled at top), specified at $t\!=\!-100M$. Time flows downwards (labeled at
left). The bottom panels show the sinusoidal dependence of the final kick
magnitude on the initial orbital phase. This animation is available at
\href{https://vijayvarma392.github.io/binaryBHexp/\#sine_kicks}{vijayvarma392.github.io/binaryBHexp/\#sine\_kicks}.
}
\label{fig:sine_kicks}
\end{figure*}
\subsection{Sinusoidal kick dependence}
As suggested above, the remnant kick is quite sensitive to the angle between
the spins close to merger. In particular, the component of the kick parallel to
the orbital angular momentum has been found to depend sinusoidally on the
orbital phase~\cite{Bruegmann-Gonzalez-Hannam-etal:2007, Gerosa:2018qay}.
Fig.~\ref{fig:sine_kicks} demonstrates this effect. All five different cases
have equal-mass BHs, with anti-parallel spins lying in the orbital plane at
$t=-100M$. Each evolution is initialized with a different orbital phase or,
equivalently, performing an overall rotation of the spins about the $z$-axis. 

As expected, the final BH kick changes dramatically with the initial orbital
phase. Even visually, the kick dependence appears to be sinusoidal. This
example demonstrates the potential of \PackageName as a tool to perform
detailed, but at the same time accessible, exploration of the phenomenology of
precessing BH mergers.

\section{Public Python implementation}
\label{sec:python_implementation}

Our package is made publicly available through the easy-to-install-and-use
Python package, \PackageName~\cite{binaryBHexp}. Our code is compatible with
both \texttt{Python~2} and \texttt{Python~3}. The latest release can be
installed from the Python Package Index using
\begin{verbatim}
    pip install binaryBHexp
\end{verbatim}
This adds a shell command called \PackageName, which can be used to generate
visualizations with invocation as simple as
\begin{verbatim}
    binaryBHexp --q 2 --chiA 0.2 0.7 -0.1 --chiB 0.2 0.6 0.1
\end{verbatim}
Such an invocation yields a running movie that the user can interact with.  By
clicking and dragging on the movie as it plays, the user can change the viewing
angle and the waveform time-series will update in real time as the viewing
angle is manipulated. The full documentation for command-line arguments is
available with the \texttt{-{}-help} flag.

As mentioned in Sec.~\ref{subsec:orbital_ang_mom}, the default setting for the
spin arrows is to be proportional to the Kerr parameter of the BH, $a$.  By
passing the optional argument
\texttt{-{}-use\_spin\_angular\_momentum\_for\_arrows} to the above command,
the spin arrows can be made proportional to the spin angular momentum of the BH
instead.

Python packages {\it NRSur7dq2}~\cite{NRSur7dq2} and {\it
surfinBH}~\cite{surfinBH} are specified as dependencies and are automatically
installed by \texttt{pip} if missing. \PackageName is hosted on GitHub at
\href{https://github.com/vijayvarma392/binaryBHexp}{github.com/vijayvarma392/binaryBHexp},
from which development versions can be installed. Continuous integration is
provided by \emph{Travis}~\cite{travis-ci}.  More details about the Python
implementation, as well as animations corresponding to the examples discussed
in this paper are available at
\href{https://vijayvarma392.github.io/binaryBHexp/}{vijayvarma392.github.io/binaryBHexp}.

\section{Conclusion}
\label{sec:conclusion}

We present a tool for visualizing mergers of precessing binary black holes.
Rather than rely on expensive numerical simulations, we base our animations on
surrogate models of numerical simulations. These are inexpensive but very
accurately reproduce numerical simulations. Therefore, we can generate
visualizations anywhere in the parameter space of the underlying surrogate
models, within a few seconds.

We make our code available through an easy-to-install-and-use python package
\PackageName~\cite{binaryBHexp}. We demonstrate the power of this tool by
generating visualizations of several well known phenomena such as: spin and
waveform modulations due to precession, orbital-hangup effect, super kicks,
sinusoidal behavior of the remnant kick, etc. This tool can be used by
researchers and students alike, to gain valuable insights into the highly
complex dynamics of precessing binary black holes.

\section*{Acknowledgements }
We thank Harald Pfeiffer for useful comments. V.V.\ is supported by the Sherman
Fairchild Foundation and NSF grants PHY--1404569, PHY--170212, and PHY--1708213
at Caltech.  D.G.\ is supported by NASA through Einstein Postdoctoral
Fellowship Grant No.\ PF6--170152 awarded by the Chandra X-ray Center, which is
operated by the Smithsonian Astrophysical Observatory for NASA under Contract
NAS8--03060.

\newcommand{\newblock}{}
\raggedright{}
\bibliographystyle{iopart_num}
\bibliography{References}

\end{document}